\documentclass[twocolumn]{aastex631}
\usepackage{mathtools}
\usepackage{amsmath}
\usepackage{CJKutf8}
\usepackage{CJK}
\usepackage{xcolor}
\usepackage{natbib}
\usepackage{longtable}

\newcommand{\Ha}{H$\alpha$}
\newcommand{\Hb}{H$\beta$}

\newcommand{\OIII}{[O\,{\sc iii}]}

\newcommand{\NII}{[N\,{\sc ii}]}

\newcommand{\reff}{$R_{\rm eff}$}
\newcommand{\ewha}{EW(H$\alpha$)}
\newcommand{\logm}{log($M/M_{\odot}$)}
\newcommand{\vmax}{1/V$_{\rm max}$}
\defcitealias{Fu18}{Paper I}
\defcitealias{Steffen21}{Paper II} 

\begin{document}
\begin{CJK}{UTF8}{gbsn}
\title{SDSS-IV MaNGA: How Galaxy Interactions Influence Active Galactic Nuclei}

\author{Joshua~L.~Steffen}
\affiliation{Department of Physics \& Astronomy, University of Iowa, Iowa City, IA 52242, USA}

\author{Hai~Fu}
\affiliation{Department of Physics \& Astronomy, University of Iowa, Iowa City, IA 52242, USA}

\author{Joel~R.~Brownstein}
\affiliation{Department of Physics and Astronomy, University of Utah, 115 S. 1400 E., Salt Lake City, UT 84112, USA}

\author{J.~M.~Comerford}
\affiliation{Center for Astrophysics and Space Astronomy, Department of Astrophysical and Planetary Sciences, University of Colorado, 389 UCB, Boulder, CO 80309-0389, USA}

\author{I.~Cruz-González}
\affiliation{Instituto de Astronomıa, Universidad Nacional Autónoma de México, A.P. 70-264, 04510, Mexico, D.F., México}

\author{Y.~Sophia~Dai~(戴昱)}
\affiliation{National Astronomical Observatories of China, Chinese Academy of Sciences, 20A Datun Road, Chaoyang District, Beijing 100012, China}

\author{Niv~Drory}
\affiliation{McDonald Observatory, The University of Texas at Austin, 1 University Station, Austin, TX 78712, USA}

\author{Arran~C.~Gross}
\affiliation{Department of Physics \& Astronomy, University of Iowa, Iowa City, IA 52242, USA}

\author{C.~Alenka~Negrete}
\affiliation{Instituto de Astronomıa, Universidad Nacional Autónoma de México, A.P. 70-264, 04510, Mexico, D.F., México}
\affiliation{CONACyT Research fellow}

\author{Renbin~Yan}
\affiliation{Department of Physics, The Chinese University of Hong Kong, Shatin, N.T., Hong Kong, China}

\begin{abstract}
We present a comparative study of active galactic nuclei (AGN) {between} galaxy pairs and isolated galaxies with the final data release of the MaNGA integral field spectroscopic survey. We build a sample of 391 kinematic galaxy pairs within the footprint of the survey and select AGN using the survey’s spectra. We use the comoving volume densities of the AGN samples to quantify the effects that tidal interactions have on the triggering of nuclear accretion. Our hypothesis is that the pair sample contains AGN that are triggered by not only stochastic accretion but also tidally induced accretion and correlated accretion. With the level of stochastically triggered AGN fixed by the control sample, we model the strength of tidally induced accretion and correlated accretion as a function of projected separation ($r_p$) and compare the model expectations with the observed volume densities of dual AGN and offset AGN (single AGN in a pair). {At $r_p$ $\sim$ 10 kpc}, we find that tidal interactions induce $\sim$30\% more AGN than stochastic fueling and cause $\sim$12\% of the offset AGN to become dual AGN because of correlations. The strength of both these effects decreases with {increasing} $r_p$. We also find that the \OIII\ luminosities of the AGN in galaxy pairs are consistent with those found in isolated galaxies, likely because stochastically fed AGN dominate even among close pairs. {Our} results illustrates that while we can detect tidally induced effects statistically, it is challenging to separate tidally induced AGN and stochastically triggered AGN in interacting galaxies.
\end{abstract}



\section{Introduction}\label{sec:intro}
\end{CJK}
The source of fueling for active galactic nuclei (AGN) has been a major topic of debate for many years. The central supermassive black hole (SMBH) requires a constant stream of accreting material to sustain the AGN; however, the angular momentum of the material beyond the galaxy's very center is too high to be accreted onto the SMBH. Some mechanism must be responsible for reducing this material's angular momentum before it may become a fuel source for the AGN. It has been long suspected that galaxy interactions may be the primary mechanism of AGN fueling as AGN were often seen within interacting galaxy systems \citep{Adams77}. Hydrodynamical simulations of galaxy mergers predict that the gravitational torques between the galaxies create {large}-scale bars which will funnel gases from the galaxies' disks towards their centers \citep{Barnes96}. These inflowing gases are predicted to trigger new star formation, dilute central metallicities, and potentially fuel the central SMBH. 

More recent observational studies have confirmed that galaxy mergers do indeed enhance central star formation {\citep{Ellison08, Kartaltepe12, Patton13, Steffen21}} and dilute central metallicities \citep{Scudder12}, but the extent to which galaxy interactions influence AGN fueling is still a matter of debate. Some studies have found no correlation between galaxy mergers and AGN {\citep{Cisternas11, Villforth17, Marian19, Silva21}} and other studies have found that AGN are statistically more likely to be in merging galaxies {\citep{Ellison11, Ellison13, Lackner14, Satyapal14, Weston17, Goulding18}}. 

These disagreements suggest that the methods for selecting galaxy pairs, control galaxies, and AGN need to be carefully considered. It has been previously observed that the AGN fraction in pairs depends strongly on the projected separation between the two galaxies and weakly depends on the stellar mass ratio between the two galaxies \citep{Ellison11}. Further, late-stage mergers that are coalescing into a single galaxies also show increased AGN fractions \citep{Ellison13}. This agrees with hydrodynamical simulations which predict that AGN activity generally increases with more advanced merger-stages but especially so as the pair coalesces into a post-merger galaxy \citep{Van-Wassenhove12, Capelo17}.

Even while an AGN may be observed in a paired galaxy, the SMBH may be fueled through stochastic processes instead of merger-induced ones. To isolate the two effects, the AGN fraction in paired galaxies needs to be compared to the AGN fraction in isolated galaxies. It is well known that the AGN fraction in galaxies depends on their redshift and stellar mass \citep{Kauffmann03}. Thus, the AGN excess, the ratio between the fraction of AGN in pairs and isolated galaxies, must be calculated between paired galaxies and control galaxies that have similar stellar masses and redshifts. 

Different AGN selection methods also yield different levels of excess AGN in pairs. \citet{Ellison11} showed that the AGN excess in paired galaxies is higher when optically selecting for composite starburst/AGN galaxies than it is for optically selecting pure AGN galaxies. It was also shown in \citet{Satyapal14} that the AGN excess nearly doubles by using infrared color cuts to select AGN in comparison to using optical selected AGN. The optical component of the AGN spectrum can be obscured by optically thick dust dredged up by the merger, thus suppressing the excess of optically selected AGN.

Along with the sample selection, the biases of the surveys' themselves need to be considered. Many of the past works using optically selected AGN have used single fiber spectroscopic surveys, like SDSS (Sloan Digital Sky Survey; \citet{Blanton17}). Surveys like SDSS have allowed for large samples of paired galaxies; however, such large surveys have some limitations. SDSS has a fiber collision limit of 55\arcsec, so that close galaxy pairs will only be found in regions between overlapping tiling plates and in {repeated} observations. SDSS also only observes targets whose brightness is greater than a r-band apparent magnitude of $r$ $<$ 17.8. These limitations create artificial biases in pair, control, and AGN samples constructed within such surveys. Single fiber surveys also have fixed angular apertures that will cover variable physical radii depending on the galaxy's redshift. Therefore, the spectra collected from the center of a high redshift galaxy may be diluted by the spectra from its disk. 

The advent of massive Integral Field Spectroscopy (IFS) surveys resolves a number of the previously mentioned issues. IFS surveys like MaNGA (Mapping Nearby Galaxies at Apache Point Observatory survey; \citet{Bundy15}) allocate a bundle of fiber optic cables to a single target, giving complete spectroscopic coverage over the whole disk of the galaxy. This allows for the construction of physical apertures based on the target's redshift instead of being limited to fixed angular apertures. Close galaxy pairs will also fall within the field-of-view of some observations, yielding spectroscopic data for both galaxies. 

In the previous paper of this series, \citet{Fu18} (hereafter; \citetalias{Fu18}), we built a sample of 105 galaxy pairs and 14 dual AGN (dAGN) in the fourteenth public data release of the MaNGA survey (MaNGA DR14). To account for MaNGA's survey biases, we compared the comoving volume densities of AGN in paired galaxies to the expected volume densities. The expected volume density of AGN are calculated from a sample of isolated control galaxies and depended on the stellar mass and redshift of the galaxies. We found that galaxy pairs hosting a single AGN (hereafter; an offset AGN) showed little evidence for an excess volume density of AGN while dAGN showed an excess volume density that increased from 3$\times$ to 6$\times$ at pair separations of 10$-$30 kpc and 1$-$10 kpc respectively. 

{The {results} of \citet{Fu18} suggested that merger events do not enhance the incidence of AGN, except in cases of simultaneous AGN. We refer to the event of simultaneous AGN in a galaxy pair as a correlated AGN. There are a number of possible sources for correlated AGN. The first is from the synchronous fueling of AGN by tidally induced gas inflows, which is predicted to occur in limited windows of projected separation \citep{Van-Wassenhove12}. The second possible source may be widespread radiative shocks from the tidally induced gas inflows. The third cause may be from AGN cross-ionization where the AGN in one of the paired galaxies is able to photoionize the gases in the neighboring galaxy. In the first case, the observed AGN are true AGN, while in the later two cases, one or both of the AGN are false detections. }

In a subsequent work, we built a more rigorous methodology for selecting pairs in MaNGA, which we used to study the radial profiles of star formation in galaxy pairs \citep{Steffen21} (hereafter; \citetalias{Steffen21}). Now, the MaNGA survey has finished its final round of observations, increasing the sample from 2618 galaxies in \citetalias{Fu18} to 10,130 galaxies. Given the improved pair selection methodology and larger galaxy sample in MaNGA, in this work we revisit the question of AGN volume density in MaNGA galaxy pairs. The larger sample size reveals an excess volume density of offset AGN at close separations which was previously unobserved in \citetalias{Fu18}. These new results allow us to separately model the effects of merger-induced AGN and correlated AGN in galaxy pairs.  


The paper is organized as follows. In Section \ref{sec:data} we discuss the MaNGA survey which we use in this study. In Section \ref{sec:pair} we discuss how we construct our pair sample, in Section \ref{sec:agn} we discuss how we select AGN, and in Section \ref{sec:control} we discuss our control sample. In Section \ref{sec:den} we study the volume densities of AGN in our pair sample in comparison with a control sample. In Section \ref{sec:oiiilum} we study the \OIII\ luminosity between AGN in our pair sample and AGN in our control sample. In Section \ref{sec:disc} we discuss the results of this work and in Section \ref{sec:conl} we summarize the results of the work. Throughout this work, we adopt the following $\Lambda$CDM cosmology; $\Omega_m$ $=$ 0.3, $\Omega_\Delta$ $=$ 0.7, and $h$ $=$ 0.7.

\section{Data: SDSS-IV MaNGA}\label{sec:data}

MaNGA (Mapping Nearby Galaxies at Apache Point Observatory) is an IFS survey which uses SDSS's 2.5 m telescope \citep{Gunn06} along with two dual-channel BOSS spectrographs \citep{Smee13}. The survey uses 17 integral field units (IFU) of sizes of 19, 37, 61, 91, and 127 fibers covering 12.5\arcsec, 17.5\arcsec, 22.5\arcsec, 27.5\arcsec, and 32.5\arcsec on the sky respectively \citep{Drory15, Law15}. MaNGA has a spectral coverage from 3600$-$10300 \AA\ with a resolution of R $\sim$ 2000 and a point spread function of 2.5\arcsec\ FWHM \citep{Bundy15}. 

With the final public data release (DR17, \citet{Abdurrouf21}) MaNGA has surveyed 10,130 galaxies which have been selected from a subset of 41,154 galaxies from the NASA-Sloan Atlas\footnote{NSA v1\_0\_1; \url{http://www.nsatlas.org}} \citep{Wake17}. The selected galaxies are chosen from a redshift range of 0.01 $<$ $z$ $<$ 0.15 and a luminosity range of -17.7 $<$ $\mathcal{M}_{\rm i}$ $<$ -24.0, where $\mathcal{M}_{\rm i}$ is the rest frame i-band absolute magnitudes within the survey's elliptical Petrosian apertures. The survey is designed to have a flat stellar mass distribution and to have a spatial coverage of 1.5 \reff\ and 2.5 \reff\ (where \reff\ is the 50\% half light radius). 

The MaNGA survey has three main subsamples; the Primary sample ($\sim$47\% of the survey), the Secondary sample ($\sim$37\% of the survey), and the Color-Enhanced sample ($\sim$16\% of the survey). The Primary sample consists of the galaxies which are covered out to 1.5 \reff\ while the Secondary sample consists of the galaxies which are covered out to 2.5 \reff. The Color-Enhanced sample is designed to target rare galaxies such as high-luminosity blue galaxies and low-luminosity red small galaxies. The Color-Enhanced sample is covered out to 1.5 \reff\ and can be combined with the Primary sample to make the Primary+ sample. In this work we will be using both the Primary+ and Secondary samples which we will refer to as the Main sample. MaNGA also contains a number of ancillary observations that were specifically designed to target galaxy mergers and AGN. We decide not to use these ancillary targets as they will bias our sample.

We construct the luminosity function for the selected galaxies to check if it follows the expected luminosity distribution of galaxies. We find that galaxies above an r-band absolute magnitude of $\mathcal{M}_{\rm r}$ $\ge$ -19 are under-sampled in our selection. We decide to remove these faint objects from our sample to avoid incompleteness.  {The preceding restrictions give us 8,585 MaNGA IFUs to work with for this study.}

\section{Pair Selection}\label{sec:pair}

Previous pair studies with MaNGA have found pairs by cross-matching MaNGA targets with galaxies in its parent catalog, the NSA catalog \citep{Pan19, Steffen21}. To utilize the volume density method {(Described in Section \ref{sec:den})}, we will compile a pair sample that is entirely contained within the MaNGA survey's 2392 arcmin$^2$ footprint. We first identify and spectroscopically classify discrete objects in the MaNGA survey's footprint. We name this catalog the {\sc mangaObj} and describe its construction fully in Steffen et al. 2022b (in preparation). 

Briefly, we first overlay the SDSS photometric catalog, {\sc photoObj}, over MaNGA's fields-of-view. The cross-matched objects from the {\sc photoObj} catalog are then visually inspected. SDSS's object deblending algorithm frequently place multiple targets on clumpy extended objects such that a single galaxy will be assigned multiple targets. We remove these over-deblended objects and then visually classified the remaining objects with MaNGA's spectroscopy. The catalog contains {11,072} galaxies that have redshifts which are similar ($z$ $<$ 0.01) to the MaNGA target galaxies or are MaNGA target galaxies themselves. {9,470 of these galaxies are within the 8,585 MaNGA IFUs that we have selected for this study.}

We use the spectral fitting code, {\sc spfit}\footnote{\url{https://github.com/fuhaiastro/spfit}}, to model the spectra in MaNGA's data-cubes \citepalias{Fu18}. {\sc spfit} is designed to simultaneously fit stellar continuum and emission lines following the Penalized Pixel-Fitting method (pPXF; \citet{Cappellari04, Cappellari17}) using the simple stellar population (SSP) models from MIUSCAT \citep{Vazdekis12}. We use the redshift of the MaNGA target, from the NASA-Sloan Atlas (NSA), as the best guess of the redshift for all objects in the MaNGA field. Objects with substantially different redshifts, $\Delta v \ge$ 2000 km s$^{-1}$, will fail to be properly modeled. {\sc spfit} provides emission line fluxes, equivalent widths, and gas kinematics calculated from the fitted emission lines and stellar kinematics and stellar masses from the stellar continuum. 

We use the galaxies with redshifts that are similar to the MaNGA target galaxy ($z$ $<$ 0.01) as candidates for our pair sample. We extract and model spectra for each object in the {\sc mangaObj} catalog with {\sc spfit} through a 1 kpc radius circular aperture centered on the galaxy's photometric center. We use the kinematics and stellar masses from {\sc spfit} in the subsequent pair selection. 

The MaNGA survey imposes a natural projected separation limit of $\sim$85 kpc across the diameter of an IFU assuming the largest IFU size, 32.5\arcsec\ and the survey's upper redshift limit, $z$ $=$ 0.15. In most cases, the MaNGA target galaxy is in the center of the IFU, so the effective projected separation limit is $\sim$42 kpc. We require that the MaNGA target galaxy is within 10\% of the radius of the IFU and that a paired galaxy be within the inscribed circle that is 1\arcsec\ from the hexagonal IFU boundary. This eliminates any geometric biases that may arise from off-center MaNGA targets or objects falling in one of the IFU's corners. {With these geometric requirements, the projected separation range of our pair sample varies from 1 kpc $-$ 36 kpc.}

We use the relative line-of-sight velocities of the galaxies to remove projected pairs from the sample. From {\sc spfit}, we have stellar and gas kinematics for the fitted spectra. We chose to use the stellar kinematics for the relative velocity cuts since galaxies hosting AGN may have large offsets due to AGN outflows. We then set a relative line-of-sight velocity of $\Delta v \le$ 500 km s$^{-1}$ to select kinematic pairs. A cut at 500 km s$^{-1}$ is subjective; however, in \citetalias{Fu18} we observed that a more or less stringent requirement on the relative line-of-sight velocity has little impact on the resulting pair sample.

The MaNGA survey's stellar mass measurements are derived from the NSA catalog. The algorithm in the NSA catalog frequently under-deblends close objects such that the stellar mass given for a MaNGA target in a pair will actually be the combined stellar mass of both galaxies. Separating this stellar mass between the two paired galaxies is not a simple task as these galaxies often overlap each other and have irregular morphologies.

As a simple attempt to resolve this problem, we use the stellar masses calculated from the inner 1 kpc radius circle from {\sc spfit} to define our mass ratios. These mass ratios are then used to partition the stellar masses from the NSA catalog between the galaxies in pairs. The mass ratio, $\mu$, is defined as;
\begin{equation}
\mu = M_{\rm t}^{SPFIT} / M_{\rm c}^{SPFIT}, 
\end{equation}
Where $M_{\rm t}^{SPFIT}$ and $M_{\rm c}^{SPFIT}$ are the stellar masses from {\sc spfit} modeled spectra extracted from a 1 kpc circular aperture for the MaNGA target and its companion respectively. This mass ratio is then used to partition the stellar mass from the NSA catalog such that the MaNGA target's stellar mass, $M_{\rm t}$, is calculated as,
\begin{equation}
M_{\rm t} = M_{\rm NSA} \, \frac{\mu}{1+\mu},
\end{equation}
and the stellar mass of the companion galaxy, $M_{\rm c}$, is then calculated as,
\begin{equation}
M_{\rm c} = M_{\rm NSA} \, \frac{1}{1+\mu}.
\end{equation}

For the remainder of the paper, we will be referring to the mass ratio, $\mu$, in logarithmic space such that log$(\mu)$ $=$ log$(M_{\rm t}^{SPFIT} / M_{\rm c}^{SPFIT})$. For the pair selection, we impose a logarithmic mass ratio cut of $|$log($\mu$)$|$ $\le$ 1.0 to remove insignificant companions from the sample. 

The pair selection process can be briefly summarized as the following. We require that:
\begin{enumerate}
	\item the pair be in either MaNGA's Primary+ or Secondary samples.
	\item the MaNGA target has a r-band absolute magnitude that is less than $\mathcal{M}_{\rm r}$ $\le$ -19.
	\item each paired observation is entirely contained within a single IFU.
	\item the MaNGA target galaxy is within 10\% of the radius of the MaNGA IFU from the IFU's center. 
	\item the galaxy pair be within the inscribed circle that is 1\arcsec\ from the IFU boundary.
	\item the relative line-of-sight velocity between the galaxies be within $\Delta v \le$ 500 km s$^{-1}$.
	\item the mass ratio between the galaxies be within $|$log($\mu$)$|$ $\le$ 1.0.
\end{enumerate}

Using the above criteria, we find 391 pairs, 35 triplets, 1 quadruplet, and 1 quintuplet {within the 8,585 MaNGA IFUs.} {If we define a major merger as paired galaxies whose mass ratio is below $|$log($\mu$)$|$ $\le$ 0.5 and a minor merger as paired galaxies whose mass ratio is between 0.5 $<$ $|$log($\mu$)$|$ $\le$ 1.0, we find 178 major mergers and 213 minor mergers among the 391 paired galaxies.} Going forward, we will focus on the galaxy pairs in order to simplify the resulting statistical analysis.
 
 \section{AGN Selection}\label{sec:agn}
\begin{figure*}
\centering
\includegraphics[width=6in]{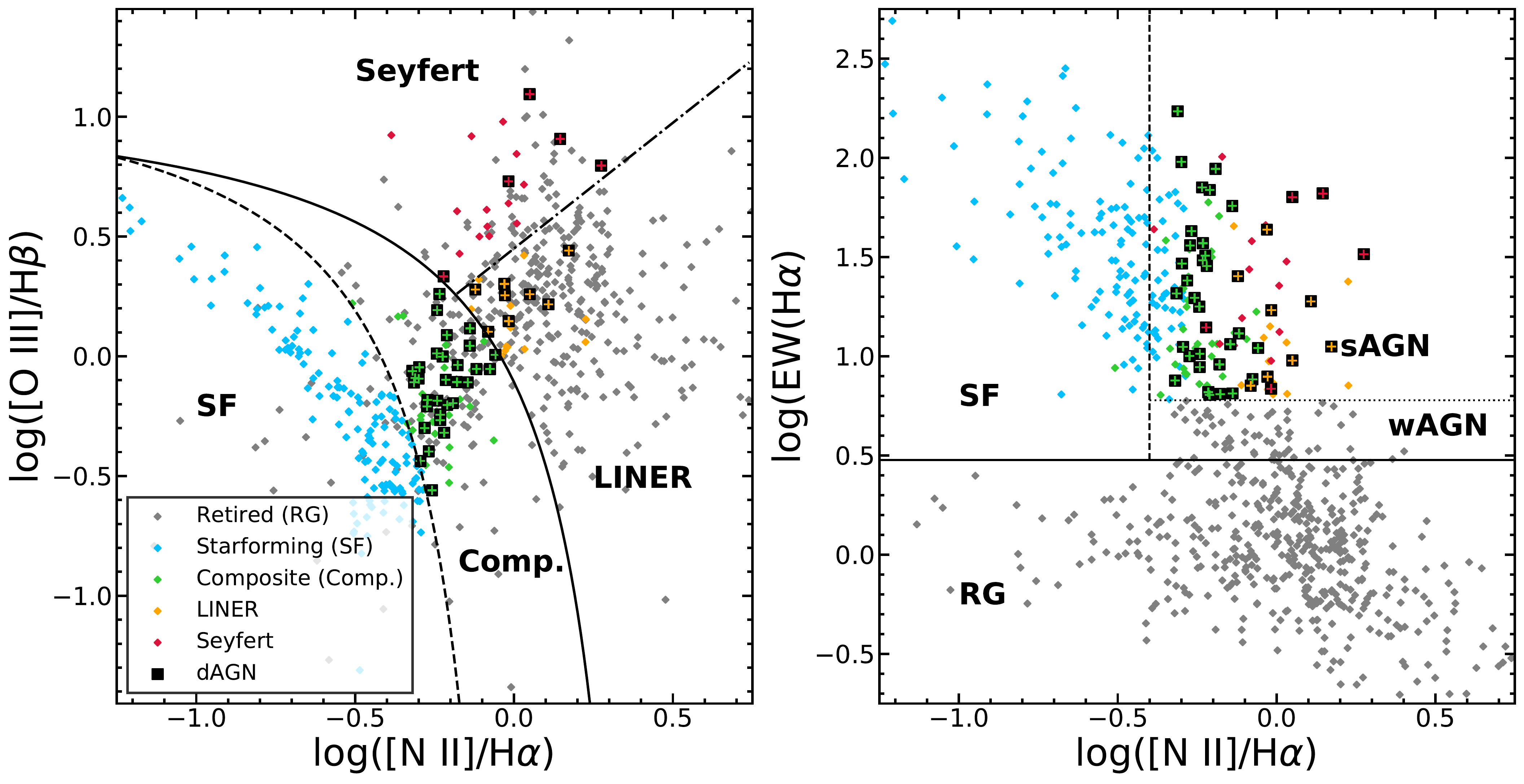}
\caption{Left: The BPT diagram of the galaxies in our pair sample. The emission line measurements are extracted from a 1 kpc radius aperture. The dashed line is the K03 line, the solid line is the K01 line, and the dot-dash line is the S07 line. The emission line classifications are color coded where Retired, Star-forming, Composite, LINER, and Seyfert are represented with grey, blue, green, orange, and red diamonds respectively. The black squares with color crosses represent the dAGN in the sample. Right: The WHAN diagram for our pair sample. The dotted horizontal line represents \ewha\ $=$ 6\AA\ and the solid horizontal line represents \ewha\ $=$ 3\AA. The \ewha\ $=$ 6\AA\ line separates weak AGN (wAGN) from strong AGN (sAGN). The color-coding of the galaxies is from their BPT classification as shown in the left panel.}
\label{fig:bpt}
\end{figure*}

We classify AGN in our sample following the BPT \citep{Baldwin81} and WHAN \citep{Cid-Fernandes11} methods. For the BPT method, we use the emission line flux ratios, log(\OIII/\Hb) and log(\NII/\Ha), extracted from a 1 kpc aperture over the centers of the galaxies. We classify AGN from star formation on the BPT diagram using the empirical separation between star-forming galaxies and AGN from \citet{Kauffmann03} (K03),
 \begin{equation}\label{eq:k03}
{\rm log}([{\rm O\,{\sc III}}]/{\rm H}\beta) = \frac{0.61}{\rm log([N\,{\sc II}]/H\alpha)-0.05} + 1.30,
\end{equation}
 the theoretical maximum starburst line from \citet{Kewley01} (K01), 
 \begin{equation}\label{eq:k01}
{\rm log}([{\rm O\,{\sc III}}]/{\rm H}\beta) = \frac{0.61}{\rm log([N\,{\sc II}]/H\alpha)-0.47} + 1.19,
\end{equation}
 and the empirical separation between Seyferts and LINERs from \citep{Schawinski07} (S07),
 \begin{equation}\label{eq:s07}
{\rm log}([{\rm O\,{\sc III}}]/{\rm H}\beta) = 1.05 {\rm log([N\,{\sc II}]/H\alpha)} + 0.45.
\end{equation}
We determine a galaxy to be on the AGN branch if it is above the K03 line. We then split the AGN branch into further categories. We classify the galaxy as composite starburst/AGN if it is between the K01 and K03 lines. We classify the galaxy as LINER if it is above the K01 line but below the S07 line. We classify the galaxy as a Seyfert if it is above both the K01 and S07 lines {(Figure \ref{fig:bpt})}. 

It has been shown that the ionization spectra from hot low-mass evolved stars (HOLMES) can create line ratios which mimic other classifications on the BPT diagram, especially LINERs \citep{Stasinska08}. \citet{Cid-Fernandes11} showed that the \ewha\ is a good tracer for these HOLMES. An \ewha\ cut of 3\AA\ is typically used to remove retired galaxies; however, we decide to use a stricter \ewha\ $<$ 6\AA\ to remove wAGN (weak AGN). We find that using the stricter \ewha\ $=$ 6\AA\ cut improves the observed volume density of AGN in the pair sample.

We also visually inspect the spectra of each of the identified galaxies for broadened emission lines. These broadened emission lines are evidence of high velocity gases near the SMBH and are a guaranteed signature of AGN activity. Such galaxies are known as broad-line AGN (BLAGN) or Type I Seyferts. If a galaxy in our sample possesses broadened emission lines, we include it in our AGN sample, regardless of its BPT or WHAN classification.

In total, we have {1045} AGN {out of the {9,470} identified galaxies, such that about 10\% of our galaxy sample hosts an AGN}. Of these, {713} are composite starburst/AGN galaxies, {97} are LINERs, {158} are Type II Seyferts, and {77} are Type I Seyferts. In the pair sample there are 105 AGN, of which 62 are composite starburst/AGN galaxies, 18 are LINERs, 17 are Type II Seyferts, and 8 are Type I Seyferts. Among the galaxy pairs with AGN, we find 25 dual AGN (dAGN) systems, which we list in Table \ref{tab:dagn}, {and we find 55 pair systems with a single AGN (hereafter offset AGN)}.

\begin{table*}
\scriptsize
\begin{center}
\caption{Dual AGN in MaNGA}
\label{tab:dagn}
\begin{tabular}{lc ccccccccc}
\hline
\hline
Plateifu & RA & DEC & Redshift & $\Delta\theta$ & $r_p$ & $\Delta v$ & $\Delta$log(M) & log(M) & BPT & log(L\OIII) \\ 
 & (deg) & (deg) &  & (arcseconds) & (kpc) & (km s$^{-1}$) & log(M$_\odot$) & log(M$_\odot$) & & log(L$_\odot$) \\ 
 (1) & (2) & (3) & (4) & (5) & (6) & (7) & (8) & (9) & (10) & (11)\\ 
\hline

7443-12703 & 229.52558 & 42.74585 & 0.04027 & \nodata & \nodata & \nodata & \nodata & 10.4 & 2 & 41.6\\ 
7443-12703 & 229.52653 & 42.74407 & 0.04027 & 6.9 & 5.5 & 89.3 & -0.1 & 10.5 & 2 & 40.8\\ 
7975-12704 & 324.58641 & 11.34867 & 0.08886 & \nodata & \nodata & \nodata & \nodata & 11.0 & 2 & 40.3\\ 
7975-12704 & 324.58655 & 11.34961 & 0.08886 & 3.5 & 5.7 & 194.8 & 0.6 & 10.3 & 2 & 40.2\\ 
8083-9101 & 50.13841 & -0.33996 & 0.03828 & \nodata & \nodata & \nodata & \nodata & 10.9 & 2 & 40.4\\ 
8083-9101 & 50.14021 & -0.33923 & 0.03828 & 7.0 & 5.3 & 127.1 & 1.0 & 9.9 & 3 & 39.6\\ 
8133-12702 & 113.51094 & 43.54360 & 0.08310 & \nodata & \nodata & \nodata & \nodata & 11.2 & 5 & \nodata \\ 
8133-12702 & 113.51438 & 43.54479 & 0.08310 & 10.0 & 15.6 & 188.7 & 0.7 & 10.5 & 2 & 38.9\\ 
8133-12704 & 114.77573 & 44.40277 & 0.13447 & \nodata & \nodata & \nodata & \nodata & 10.9 & 3 & 40.1\\ 
8133-12704 & 114.77430 & 44.40287 & 0.13447 & 3.7 & 8.8 & 67.8 & -0.2 & 11.1 & 3 & 40.1\\ 
8332-12704 & 209.16355 & 43.58561 & 0.10309 & \nodata & \nodata & \nodata & \nodata & 11.0 & 2 & 40.7\\ 
8332-12704 & 209.16066 & 43.58578 & 0.10309 & 7.6 & 14.4 & 85.2 & 0.8 & 10.2 & 2 & 40.9\\ 
8612-12705 & 255.10152 & 38.35170 & 0.03579 & \nodata & \nodata & \nodata & \nodata & 10.5 & 2 & 40.1\\ 
8612-12705 & 255.10322 & 38.35430 & 0.03579 & 10.5 & 7.5 & 64.9 & 0.5 & 10.0 & 2 & 40.0\\ 
8652-9102 & 331.64618 & 0.05643 & 0.04614 & \nodata & \nodata & \nodata & \nodata & 10.8 & 3 & 40.8\\ 
8652-9102 & 331.64572 & 0.05766 & 0.04614 & 4.7 & 4.3 & 23.4 & 0.3 & 10.6 & 2 & 40.1\\ 
9039-9101 & 229.00036 & 34.35688 & 0.12530 & \nodata & \nodata & \nodata & \nodata & 10.8 & 2 & 40.1\\ 
9039-9101 & 229.00243 & 34.35531 & 0.12530 & 8.4 & 18.8 & 198.2 & -0.2 & 11.0 & 5 & \nodata \\ 
9088-9102 & 242.47148 & 26.62545 & 0.07786 & \nodata & \nodata & \nodata & \nodata & 11.0 & 2 & 40.1\\ 
9088-9102 & 242.47287 & 26.62491 & 0.07786 & 4.9 & 7.2 & 89.4 & -0.1 & 11.0 & 2 & 40.0\\ 
9094-1902 & 239.71019 & 27.38998 & 0.09353 & \nodata & \nodata & \nodata & \nodata & 10.7 & 2 & 41.1\\ 
9094-1902 & 239.71107 & 27.39142 & 0.09353 & 5.9 & 10.2 & 214.2 & 0.1 & 10.5 & 2 & 39.9\\ 
9504-9102 & 121.70124 & 28.42148 & 0.14231 & \nodata & \nodata & \nodata & \nodata & 10.7 & 2 & 40.5\\ 
9504-9102 & 121.69901 & 28.42025 & 0.14231 & 8.3 & 20.8 & 89.3 & -0.3 & 11.0 & 3 & 40.3\\ 
9892-6102 & 247.47037 & 24.44400 & 0.03787 & \nodata & \nodata & \nodata & \nodata & 10.4 & 5 & \nodata \\ 
9892-6102 & 247.46998 & 24.44366 & 0.03787 & 1.8 & 1.3 & 69.1 & 0.0 & 10.4 & 5 & \nodata \\ 
9892-12702 & 247.81469 & 23.88264 & 0.05918 & \nodata & \nodata & \nodata & \nodata & 10.6 & 4 & 41.2\\ 
9892-12702 & 247.81508 & 23.88313 & 0.05918 & 2.2 & 2.5 & 35.7 & 0.5 & 10.1 & 3 & 40.9\\ 
10218-12703 & 118.63425 & 16.80972 & 0.04609 & \nodata & \nodata & \nodata & \nodata & 10.8 & 5 & \nodata \\ 
10218-12703 & 118.63312 & 16.80600 & 0.04609 & 13.9 & 12.6 & 207.5 & 0.3 & 10.5 & 3 & 41.0\\ 
10518-12704 & 152.68067 & 6.20040 & 0.09775 & \nodata & \nodata & \nodata & \nodata & 11.0 & 5 & \nodata \\ 
10518-12704 & 152.67895 & 6.19949 & 0.09775 & 7.0 & 12.6 & 29.0 & 0.8 & 10.3 & 2 & 39.5\\ 
10837-9102 & 159.34846 & 2.31265 & 0.04021 & \nodata & \nodata & \nodata & \nodata & 10.8 & 3 & 40.3\\ 
10837-9102 & 159.34907 & 2.31096 & 0.04021 & 6.5 & 5.2 & 39.7 & 0.9 & 9.9 & 2 & 41.3\\ 
11867-12704 & 136.00426 & 1.45809 & 0.05338 & \nodata & \nodata & \nodata & \nodata & 10.5 & 2 & 41.0\\ 
11867-12704 & 136.00436 & 1.45942 & 0.05338 & 4.8 & 5.0 & 10.6 & 0.9 & 9.7 & 2 & 39.6\\ 
11944-12701 & 241.89255 & 36.48404 & 0.02983 & \nodata & \nodata & \nodata & \nodata & 10.3 & 4 & 40.3\\ 
11944-12701 & 241.89361 & 36.48445 & 0.02983 & 3.4 & 2.0 & 15.3 & 0.5 & 9.8 & 2 & 39.0\\ 
11980-12702 & 253.74259 & 22.14845 & 0.03544 & \nodata & \nodata & \nodata & \nodata & 10.5 & 2 & 39.8\\ 
11980-12702 & 253.74197 & 22.14883 & 0.03544 & 2.5 & 1.7 & 39.2 & 0.5 & 10.1 & 2 & 39.3\\ 
11984-3701 & 256.59261 & 21.40617 & 0.03096 & \nodata & \nodata & \nodata & \nodata & 10.1 & 2 & 40.1\\ 
11984-3701 & 256.59269 & 21.40522 & 0.03096 & 3.4 & 2.1 & 24.1 & 0.6 & 9.6 & 2 & 39.0\\ 
12080-6104 & 31.27009 & -0.71173 & 0.04319 & \nodata & \nodata & \nodata & \nodata & 10.1 & 2 & 39.6\\ 
12080-6104 & 31.26918 & -0.71187 & 0.04319 & 3.3 & 2.8 & 4.8 & 0.7 & 9.4 & 2 & 38.5\\ 
12092-3701 & 13.91369 & 14.77361 & 0.04038 & \nodata & \nodata & \nodata & \nodata & 9.9 & 2 & 39.5\\ 
12092-3701 & 13.91298 & 14.77390 & 0.04038 & 2.7 & 2.1 & 19.0 & 0.4 & 9.5 & 2 & 38.7\\ 
12512-12702 & 146.37349 & -0.36520 & 0.05150 & \nodata & \nodata & \nodata & \nodata & 10.4 & 4 & 41.5\\ 
12512-12702 & 146.37380 & -0.36844 & 0.05150 & 11.7 & 11.8 & 364.9 & -0.1 & 10.5 & 2 & 40.3\\ 
12518-3703 & 159.60443 & -0.39268 & 0.09630 & \nodata & \nodata & \nodata & \nodata & 11.1 & 4 & 42.0\\ 
12518-3703 & 159.60487 & -0.39198 & 0.09630 & 3.0 & 5.3 & 13.6 & 0.4 & 10.7 & 4 & 40.5\\ 
\hline

\end{tabular}
\end{center}

\tablecomments{
{(1) MaNGA plate-IFU number. (2)-(3) Right ascension and declination in degrees. (4) Redshift from spectra extracted through a 1 kpc radius aperture. (5)-(6) Projected separation in arcseconds and kiloparsecs. (7) line-of-sight velocity offset. (8) Logarithmic stellar mass ratio. (9) Stellar mass of the galaxy calculated by splitting the stellar masses from the NSA catalog between the two components with the mass ratio. (10) BPT classification code: 0-retired, 1-star forming, 2-composite starburst/AGN, 3-LINER, 4-type I Seyfert, 5-type II Seyfert. (11) \OIII\ luminosity in units of L$_\odot$. \OIII\ luminosities are not computed for BLAGN.}}
\end{table*}

\section{Control Sample}\label{sec:control}
\begin{figure*}
\centering
\includegraphics[width=6in]{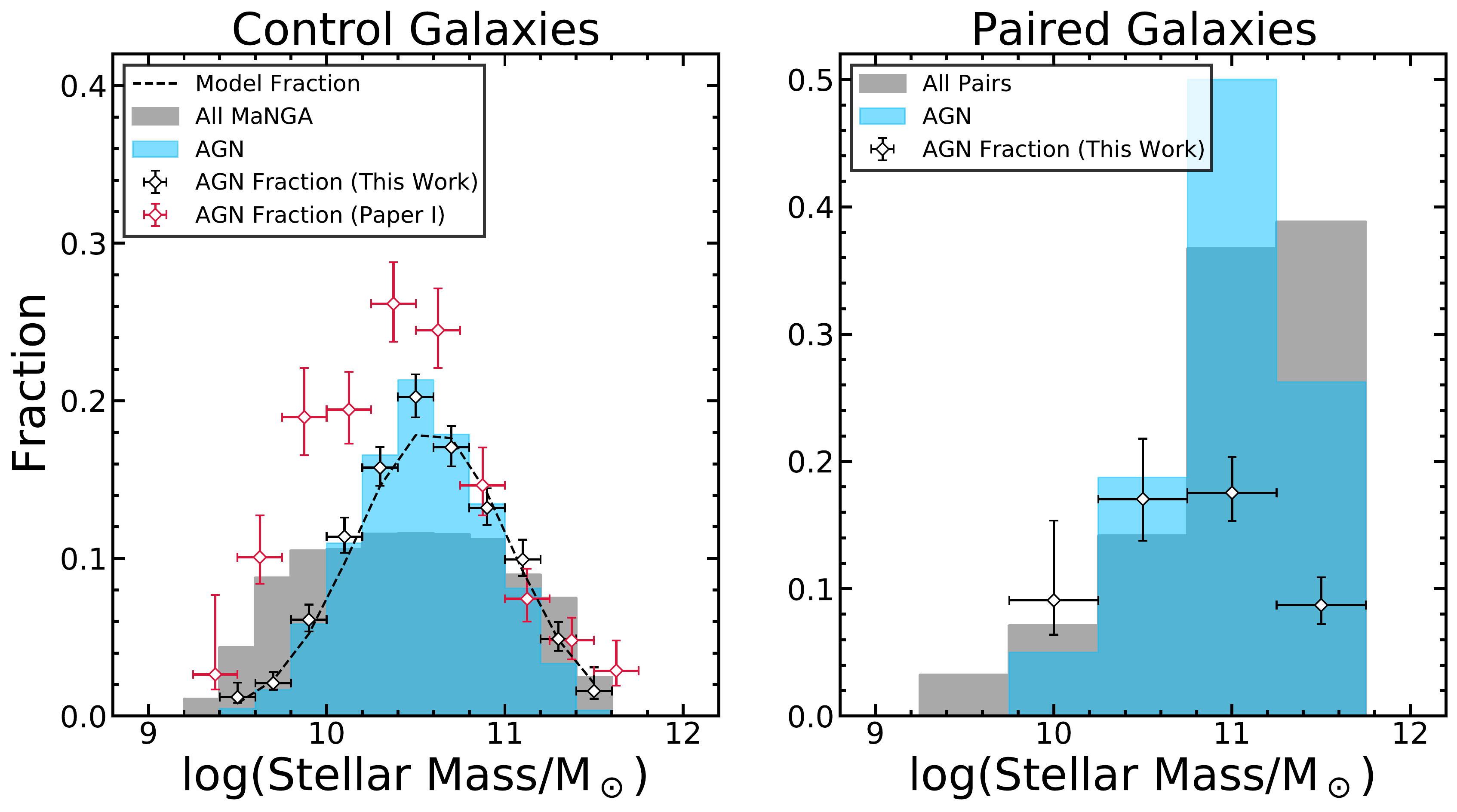}
\caption{{Left:} The stellar mass biases of the AGN sample. The grey histogram shows the stellar mass distribution of the whole MaNGA sample while the blue histogram shows the stellar mass distribution of the MaNGA galaxies hosting an AGN. The {open} black diamonds represents the AGN fraction from this work and the {open} red diamonds represent the AGN fraction found in \citetalias{Fu18}. The horizontal error bars represent the bin size and the vertical error bars represent the binomial errors of each bin. The black dashed line shows the modeled AGN fraction from this work. We see that AGN are most prevalent at stellar masses of log(M/M$_{\odot}$) $=$ 10.5. {Right: The same as the Left panel, except for the paired galaxies in our sample.}}
\label{fig:agn_frac}
\end{figure*}

Since we want to isolate merger induced effects in galaxy pairs, we need a sample of control AGN in isolated galaxies to set a baseline. Our control sample consists of 7811 isolated galaxies that have no companions within the MaNGA IFU with a line-of-sight velocity separation of $\Delta$v $\le$ 2000 km s$^{-1}$. Within these control galaxies there are 872 AGN of which; 613 are composite starburst/AGN galaxies, 70 are LINERs, 126 are narrow-line Seyferts, and 63 are BLAGN. 

In order to study AGN rates between controls and pairs, we need to control for other factors which may influence stochastic AGN rates. It has been shown in previous works that the rate of AGN is related to the stellar mass and redshift of the host galaxy. The MaNGA survey has a limited redshift range, {therefore, we do not} expect that the AGN fraction in MaNGA will have a significant dependence on redshift. We represent the {model} AGN fraction, $f^{\rm mod}_{\rm AGN}$, as a function of stellar mass, $M$. We show the AGN fraction as a function of the stellar mass for the control and pair samples in Figure \ref{fig:agn_frac}. We find that the AGN fraction as a function of the stellar mass can be fit well with a log-normal function.
\begin{equation}\label{eq:mod}
f^{\rm mod}_{\rm AGN}(M, z) = f_0 \, {\rm exp}\left[ -\frac{1}{2} \left( \frac{{\rm log}\,M/M_\odot - b}{\sigma}\right)^2\right] (1+z)^4.
\end{equation}
Even though the AGN fraction will have no significant evolution in our redshift range, we include the redshift term, $z$, in our model. We fix the power-law index to 4 so that it is consistent with the AGN luminosity function for galaxies with $z \le 1$ \citep{Ueda03}. We then fit the parameters of the stellar mass's log-normal function and find that $f_0 = 0.12$, $b = 10.56$, and $\sigma = 0.44$. 

We previously calculated the AGN fraction in \citetalias{Fu18} and found the following fit parameters, $f_0 = 0.22$, $b = 10.57$, and $\sigma = 0.54$. The amplitudes of the AGN fraction are substantially different between the works which is largely due to the stricter \ewha\ cuts that we implement in this work. If we recalculate the AGN fraction while lowering our \ewha\ cut to 3\AA, we find that $f_0 = 0.19$. This value is consistent with our previously calculated value from \citetalias{Fu18}. We can now use this model of the AGN fraction (assuming an \ewha\ cut of 6\AA) to estimate the expected volume densities of AGN in our pair sample. 

{This log-normal distribution for the stellar masses of AGN in MaNGA was previously seen in \citetalias{Fu18} and \citet{Sanchez22}. The AGN fraction peaks around \logm\ $=$ 10.5$-$11.0 because of the selection biases imposed by the BPT+WHAN selection criteria (Section \ref{sec:agn}). AGN become rarer in lower mass galaxies because they host less massive black holes. If these black holes accrete material with the same Eddington ratio as the more massive galaxies, they will be less luminous than the black holes in the higher mass galaxies. These less luminous AGN will have a smaller ionized region which may be diluted in our 1 kpc radial aperture so that they may be missed by our BPT analysis. The AGN fraction may decline again at high stellar masses because high mass galaxies tend to be quiescent and tend to have lower gas fractions. A gas-poor galaxy hosting an AGN may be missed by BPT analysis since they will not produce strong emission lines. These selection biases further emphasize the importance of the control sample in our analysis.}

\section{Volume Densities of AGN in Pairs}\label{sec:den}

\subsection{Volume Weights}\label{sec:weights}

The volume of the MaNGA survey depends on the luminosity of the galaxy and the subsample that the galaxy is in (i.e. Primary+ or Secondary). This is a consequence of the MaNGA survey's target strategy which results in the galaxies lying on two distinct bands in redshift and luminosity space. One band represents the Primary+ galaxies and the other represents the Secondary galaxies. This is important to consider when studying population statistics in the MaNGA survey. 

A volume limited sample can be recovered by using the \vmax\ weights method \citep{Schmidt68} which we used in \citetalias{Fu18}. In the MaNGA survey, the weight of a galaxy is defined in \citet{Wake17} as;
\begin{equation}\label{eq:weight}
W_j \equiv \frac{N_{\rm tiled}}{N_{\rm obs}}\frac{10^6 \, {\rm Mpc}^3}{V_{\rm tiled}\left[z_{\rm min}\left(\mathcal{M}_j\right), z_{\rm max}\left(\mathcal{M}_j\right)\right]},
\end{equation}
where $N_{\rm tiled}$ is the total number of tiling plates from MaNGA's targeting catalog which covers an area of 7362 deg$^2$ with 1800 plates, $N_{\rm obs}$ is the number of tiling plates that were used to build the final MaNGA sample, and $V_{\rm tiled}$ is the volume of the tiling plates. This tiling volume is dependent on the minimum and maximum redshifts, $z_{min}$ and $z_{max}$, at a given luminosity, $\mathcal{M}_j$. 

The 1800 targeting plates are for MaNGA's targeting catalog and the final data release of the survey only observed 609 of those targeting plates. To account for this, we recalculate the \vmax\ weights using a value of $N_{\rm tiled}$ that is calculated for the 609 plates that are actually observed by the end of the survey following the methodology from \citetalias{Fu18}.

\subsection{The Expected and Observed Volume Density}

We are now ready to compare the rate of AGN in galaxy pairs with those in isolated galaxies. At a first glance, this could be answered by comparing the fraction of AGN in pairs to the fraction of AGN in controls; however, this does not account for the mass dependence of the AGN duty cycle. The MaNGA pair sample is biased towards high stellar masses \citepalias{Fu18} and the AGN fraction peaks around \logm $=$ 10.5. Further the MaNGA sample was designed to achieve a flat stellar mass distribution so high mass galaxies are over-sampled and low mass galaxies are under-sampled. This means that a simple AGN fraction in a sample are under the influence of several factors. We therefore use the \vmax\ weights described in Section \ref{sec:weights} to calculate the volume density of AGN in the samples and fold in the mass-dependancy of stochastic AGN when calculating the expected volume density of AGN. 

The observed volume density for a set of galaxies, $n_{\rm obs}$, can be calculated as,
 \begin{equation}\label{eq:obs}
n_{\rm obs} = \sum^{N}_{j=1} W_j,
\end{equation}
where $N$ is the number of galaxies in each set and $W_j$ is the \vmax weight of the MaNGA target. 

Next, we compare the observed AGN volume density to what is predicted by the random pairing of stochastically induced AGN.  We define three subsamples. First, there is the offset AGN sample in which only one of the two components in a pair is an AGN. Second, there is the dual AGN sample in which both pair components are AGN. Third, there is the offset+dual AGN sample which contains both the offset and dual AGN samples. 

We use the AGN fraction from the isolated sample to establish the expected baseline AGN volume density. The expected volume density of dAGN from stochastically induced processes is the product of the modeled AGN fraction for both galaxies and the \vmax\ weight,
\begin{equation}\label{eq:dagn}
n_{\rm dagn} = \sum^{N_{\rm pair}}_{j=1} \,  W_j \, f_{\rm agn}^t \, f_{\rm agn}^c,
\end{equation}
where $f_{\rm agn}^t$ and $f_{\rm agn}^c$ are the stochastic AGN probabilities from the model in Equation \ref{eq:mod}, evaluated using the stellar mass and redshift of the MaNGA target and chosen companion respectively. The expected volume density of the offset AGN sample is,
\begin{equation}\begin{aligned}\label{eq:oagn}
n_{\rm oagn} = \sum^{N_{\rm pair}}_{j=1} \, W_j \, f_{\rm agn}^t\, [1 - f_{\rm agn}^c]  + \sum^{N_{\rm pair}}_{j=1} \, W_j \, f_{\rm agn}^c \, [1-f_{\rm agn}^t].
\end{aligned}\end{equation}
Finally, the expected volume density of the offset+dual AGN sample is just the sum of Equations \ref{eq:dagn} and \ref{eq:oagn}, which is equal to the following,
\begin{equation}\begin{aligned}\label{eq:odagn}
n_{\rm o+dagn} = \sum^{N_{\rm pair}}_{j=1} \, W_j \, f_{\rm agn}^t + \sum^{N_{\rm pair}}_{j=1} \, W_j \, f_{\rm agn}^c - \sum^{N_{\rm pair}}_{j=1} \, W_j \, f_{\rm agn}^t \, f_{\rm agn}^c.
\end{aligned}\end{equation}

\begin{figure*}
\centering
\includegraphics[width=7in]{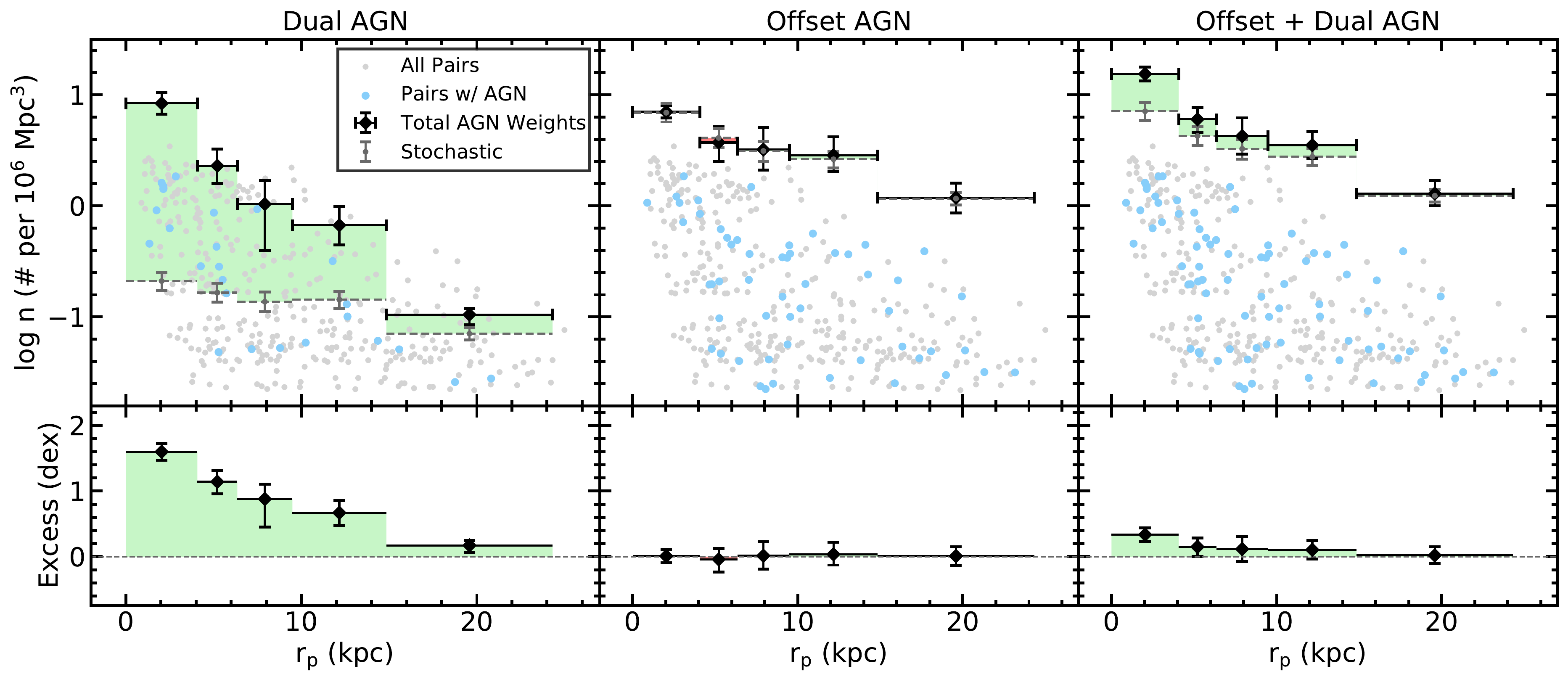}
\caption{Observed versus expected AGN volumes densities as a function of projected separation in our MaNGA pair sample. In the top panels, the grey circles represent the individual \vmax\ weights {for the whole pair sample} where the {blue} circles represent pairs with AGN. The black diamonds represent the observed volume densities, from Equation \ref{eq:obs} and the dark grey circles represent the expected volume densities (from Equations \ref{eq:dagn}, \ref{eq:oagn}, and \ref{eq:odagn} for the dual, offset, and offset+dual subsamples respectively). The horizontal error bars represent the bin size {in which the observed and expected volume densities are calculated} and the vertical error bars represent the 1$\sigma$ confidence interval. The filled color between the observed and expected volume densities represent where the observed volume density is greater than (green) or less than (red) the expected volume density. The bottom panels show the excess of observed AGN by taking the logarithmic ratio of the observed and expected volume densities. }
\label{fig:vol_den_sep}
\end{figure*}


{In Figure \ref{fig:vol_den_sep}, we present the observed (Equation \ref{eq:obs}) and expected (Equations \ref{eq:dagn}, \ref{eq:oagn}, and \ref{eq:odagn}) volume densities as a function of the projected separation between the pairs.} We split the \vmax\ weights into projected separation bins using an adaptive binning method, since the distribution of paired galaxies is not uniform across the projected separation space. In this method, we give a desired number of separation bins to split the data between. The bin edges are then defined to evenly split the sample between the bins. In Figure \ref{fig:vol_den_sep}, we split the sample of paired galaxies into five separate bins and we exclude bins which possess one or fewer AGN. By doing this, the volume densities of each projected separation bin will be calculated from the summation of a uniform number of objects.

We calculate 1$\sigma$ confidence intervals for the observed and expected volume densities using the bootstrap resampling method. We also added the error (0.01 dex) from the best-fit AGN fraction model from Equation \ref{eq:mod} to the uncertainty of the expected volume densities. 

In the left panel of Figure \ref{fig:vol_den_sep}, we see that there is a strong excess of dAGN in our pair sample. This excess is greatest at close separations, 1.6 dex (39.8$\times$) below 4 kpc, and falls with wider separations, 0.2 dex (1.6$\times$) at 20 kpc. In the middle panel of Figure \ref{fig:vol_den_sep}, the offset AGN sample shows no excess or deficit of AGN. In the right panel of Figure \ref{fig:vol_den_sep}, the offset+dual AGN sample shows an AGN excess of 0.4 dex  (2.5$\times$) below 4 kpc {which falls to zero between 4$-$20 kpc.} 

\subsection{A Qualitative Model for the AGN Excess}

\begin{figure*}
\centering
\includegraphics[width=7in]{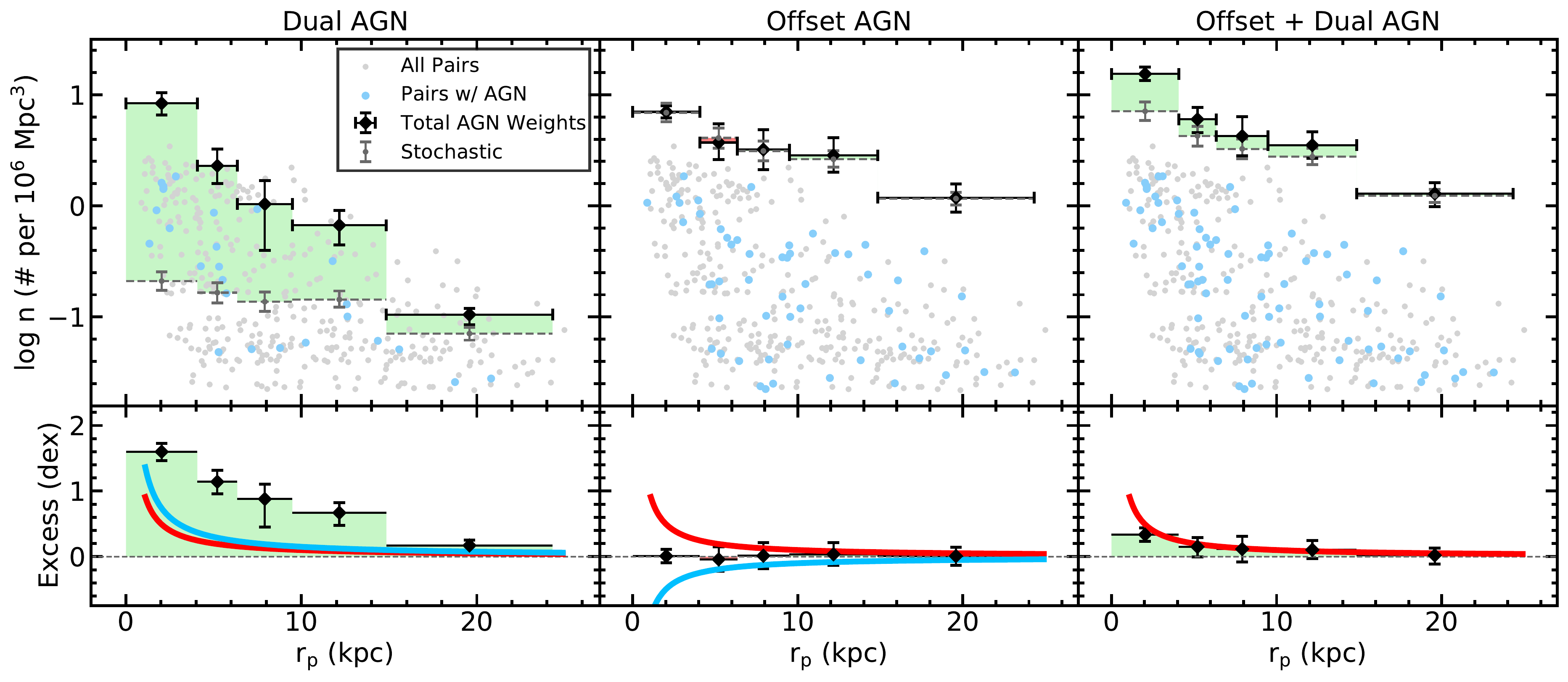}
\caption{Same as Figure \ref{fig:vol_den_sep} except that our qualitative models of the effect of correlated AGN activity (blue) and merger-enhanced AGN activity (red) are shown. {The red and blue lines are visual representations of how the two effects should affect the {resultant} volume density.}}
\label{fig:vol_den_model}
\end{figure*}

In Figure \ref{fig:vol_den_sep}, we observe a strong excess of dAGN, a mild excess of offset+dual AGN, and no excess of offset AGN. The absence of excess AGN in the offset AGN sample initially indicates that there are no merger induced AGN fueling in paired galaxies. The strong excess of dAGN then may be entirely due to correlated AGN. Correlated AGN refer to the synchronous fueling of AGN in the centers of both paired galaxies. Hydrodynamical simulations predict that the majority of AGN activity in galaxy pairs is not synchronous; however, a brief period of correlated activity occurs just prior to the coalescence of the two galaxies \citep{Van-Wassenhove12, Capelo17}. 

This model of no merger-enhanced {offset} AGN and correlated dAGN would seem to work if it wasn't for the excess of AGN in the offset+dual AGN sample. The offset+dual AGN sample should be independent of correlated AGN, so another factor is required to explain the excess observed there. Further, if there was no merger-enhanced AGN and correlated AGN, we would expect to see a deficit of offset AGN as these are being converted into dAGN by correlation. 

The reasons above lead us to conclude that a combination of merger-enhanced AGN and correlated AGN activity need to be added to the predicted volume density of AGN in offset, dual, and offset+dual AGN (Equations \ref{eq:oagn}, \ref{eq:dagn}, and \ref{eq:odagn}). We qualitatively show the affect that merger-enhanced AGN and correlated AGN would have on our excess AGN in Figure \ref{fig:vol_den_model}. The presence of merger-enhanced AGN (shown with the red line in Figure \ref{fig:vol_den_model}) would increase the volume density of AGN in all three subsamples and should be separation dependent with close galaxy pairs featuring the most merger-induced AGN. Correlated AGN (shown with the blue line in Figure \ref{fig:vol_den_model}) will increase the volume density of dAGN, while also decreasing the volume density of offset AGN. Since simulations predict that this effect is greatest at late merger stages, the effect of correlated AGN will also be separation dependent with the most correlation occurring at close separations. 

With this model we can now qualitatively explain the excess of AGN in the three subsamples. In the dAGN sample, the strong excess of AGN is due to the combination of correlated AGN and merger-enhanced AGN. In the offset AGN sample, merger-enhanced AGN is contributing excess AGN, but AGN correlation is also removing excess AGN such that we observed no excess AGN over what is predicted by the stochastic fueling of AGN. In the offset+dual AGN sample, the excess of AGN is due to the effect of merger-enhanced AGN. 

\subsection{Correlated AGN Activity and Merger Enhanced AGN Fueling}\label{sec:fit}

\begin{figure*}
\centering
\includegraphics[width=7in]{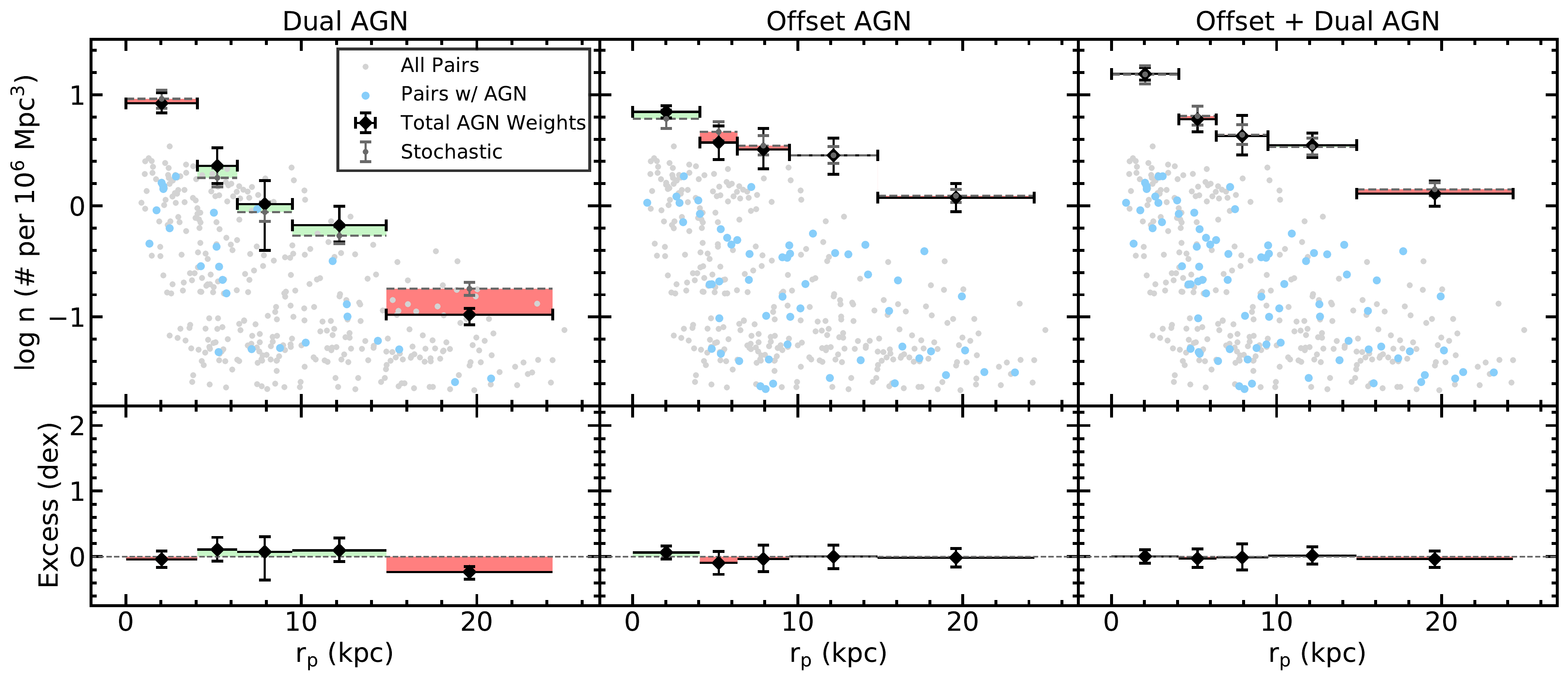}
\caption{Same as Figure \ref{fig:vol_den_sep} except that Equations \ref{eq:oagn2}, \ref{eq:dagn2} and \ref{eq:odagn2} are being used to calculate the expected volume density which account for stochastic fueling, merger-enhanced fueling, and correlated activity. }
\label{fig:vol_den_fit}
\end{figure*}

In the previous section, {it was found} that stochastic AGN fueling is not sufficient to explain the volume density of AGN in pairs that we observe. Indeed, we expect merger-enhanced fueling and correlated AGN activity to contribute to the observed volume density of AGN. In this section we redefine Equations \ref{eq:dagn}, \ref{eq:oagn}, and \ref{eq:odagn} to account for these additional parameters. We use the discrepancies between the observed and expected AGN volume densities to quantify the effects of the merger-enhanced AGN and correlated AGN.

The merger-enhanced fueling parameter will boost the original stochastic AGN rate, $f_{\rm agn}$, by a separation dependent term, $1+\lambda$, such that,
\begin{equation}
f_{\rm agn}^\prime\left(M, z\right) = \left(1+\lambda\right) \, f_{\rm agn}\left(M, z\right),
\end{equation}
where $f_{\rm agn}^\prime$ is the new merger-enhanced AGN probability. If $\lambda$ is non-zero and positive, it will increase the number of expected AGN in all three AGN-pair samples, the offset AGN, offset+dual AGN and the dAGN samples. 

In \citetalias{Fu18}, we explored the rate of correlated AGN in the MaNGA. This effect mathematically removes AGN from the offset AGN sample and adds them to the dAGN sample. To account for this, we subtract some fraction, $\xi$, from the offset AGN sample in Equation \ref{eq:oagn} such that,
\begin{equation}\begin{aligned}\label{eq:oagn2}
n_{\rm oagn}^\prime =  \sum^{N_{\rm pair}}_{j=1} \, (1-\xi) \, W_j \, f_{\rm agn}^{\prime\, t}\, [1 - f_{\rm agn}^{\prime\, c}]  \\
+ \sum^{N_{\rm pair}}_{j=1} \, (1-\xi) \, W_j \, f_{\rm agn}^{\prime\, c} \, [1-f_{\rm agn}^{\prime\, t}].
\end{aligned}\end{equation}
and allocated them to the dAGN sample in Equation \ref{eq:dagn} such that,
\begin{equation}\begin{aligned}\label{eq:dagn2}
n_{\rm dagn}^\prime = \sum^{N_{\rm pair}}_{j=1} \,  W_j \, f_{\rm agn}^{\prime\, t} \, f_{\rm agn}^{\prime\, c} +  \sum^{N_{\rm pair}}_{j=1} \, \xi \, W_j \, f_{\rm agn}^{\prime\, t} \, [1 - f_{\rm agn}^{\prime\, c}]  \\
+ \sum^{N_{\rm pair}}_{j=1} \, \xi \, W_j \, f_{\rm agn}^{\prime\, c} \, [1-f_{\rm agn}^{\prime\, t}].
\end{aligned}\end{equation}
The offset+dual AGN category in Equation \ref{eq:odagn} remains unaffected by $\xi$ as the sample does not change if an offset AGN is made into a dAGN by correlated activity, but it is still affected by merger-enhanced fueling such that,
\begin{equation}\begin{aligned}\label{eq:odagn2}
n_{\rm o+dagn}^\prime = \sum^{N_{\rm pair}}_{j=1} \, W_j \, f_{\rm agn}^{\prime\, t} + \sum^{N_{\rm pair}}_{j=1} \, W_j \, f_{\rm agn}^{\prime\, c} - \sum^{N_{\rm pair}}_{j=1} \, W_j \, f_{\rm agn}^{\prime\, t} \, f_{\rm agn}^{\prime\, c}.
\end{aligned}\end{equation}

We can now solve for the newly introduced terms, $\lambda$ and $\xi$ by matching the expected AGN volume density to the observed AGN volume density. The merger enhanced AGN will be separation dependent since closer pairs tend to be in more advanced merger stages. To account for this, we will model $\lambda$ as an inverse power law such that,
\begin{equation}\label{eq:lam}
\lambda(r_{\rm p}) = \lambda_{10} \, \left(\frac{r_{\rm p}}{\rm 10 \, kpc}\right)^{-1},
\end{equation}
where $\lambda_{10}$ is the normalization factor for $\lambda$. The correlated AGN will also be separation dependent, so the rate of correlated AGN will also be represented as an inverse power-law such that,
\begin{equation}\label{eq:xi}
\xi(r_{\rm p}) = \xi_{10} \, \left(\frac{r_{\rm p}}{\rm 10 \, kpc}\right)^{-1},
\end{equation}
where $\xi_{10}$ is the normalization factor for $\xi$. 

We use a minimization routine to fit for the unknown parameters of the inverse power-laws for both $\lambda$ and $\xi$ simultaneously. $\lambda$ is allowed to be any positive number since it functions as a multiplier to $f_{\rm agn}$. $\xi$ must be between 0 and 1 since it represents the percentage of offset AGN that are being turned into dAGN. We find that $\lambda_{10}$ $=$ 0.273 and $\xi_{10}$ $=$ 0.115 with a reduced $\chi^2$ $=$ 0.29. This solution means that at 10 kpc, the rate of stochastic fueling is being enhanced by a factor of 1.27$\times$ {(i.e. tidal interactions induce 30\% more AGN than stochastic fueling)} and 11.5\% of offset AGN are being converted into dAGN through correlated activity. {We recreate Figure \ref{fig:vol_den_sep} using Equations \ref{eq:oagn2}, \ref{eq:dagn2}, and \ref{eq:odagn2} and using the fitted values of $\lambda(r_p)$ and $\xi(r_p)$ in Figure \ref{fig:vol_den_fit}. Our updated model of the expected AGN volume density now closely follows the observed AGN volume density.}

\subsection{Mass Ratio Dependent Duty Cycle of AGN in Pairs}

We have assumed that the volume density of AGN is primarily a function of the projected separation; however, it may depend on other merger parameters. In \citetalias{Steffen21}, we observed that merger-induced star formation was dependent on both the projected separation and the mass ratio. Therefore it is reasonable to assume that the volume density of AGN may also have a dependency on the mass ratio.

\begin{figure*}
\centering
\includegraphics[width=7in]{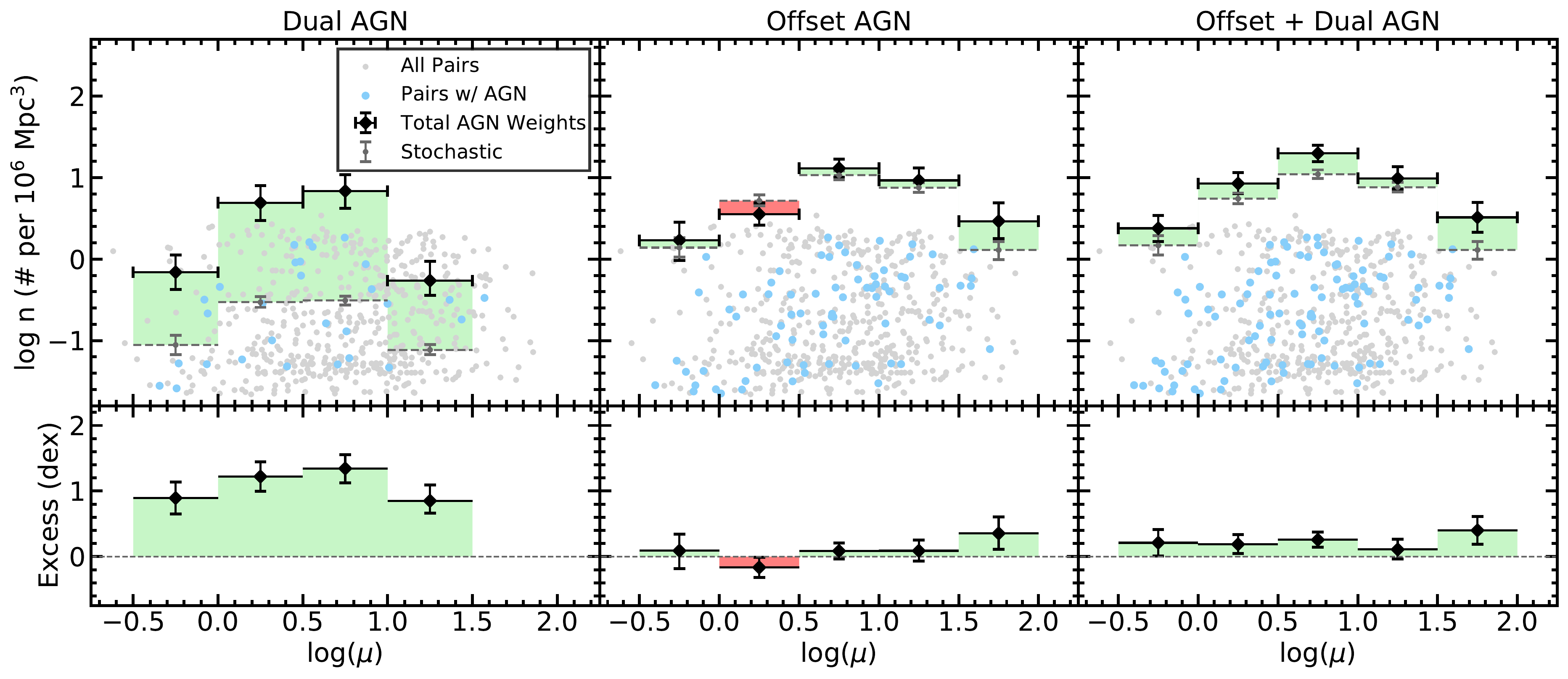}
\caption{Same as Figure \ref{fig:vol_den_sep} but calculated the observed and expected volume densities as a function of the mass ratio. }
\label{fig:vol_den_dm}
\end{figure*}

We rebuild Figure \ref{fig:vol_den_sep} but using the mass ratio, {log($\mu$)},  between the pairs to bin the volume densities in Figure \ref{fig:vol_den_dm}. For this figure, we expand the allowed mass ratio cut for the pair sample to {$|$log($\mu$)$|$} $\le$ 2.0. {The mass ratios given in Figure \ref{fig:vol_den_sep} is not taken as the absolute value of the mass ratio so that we may separately study how the more massive and less massive galaxy in a pair responds to the merger event.} In the offset AGN and offset+dual AGN samples, we see a roughly flat enhancement to the volume density of AGN across the mass ratio range of about 0.2 dex ($\sim$1.6$\times$). In the dAGN sample, the excess volume density peaks between {log($\mu$)} $=$ 0.0$-$1.0 by 1.3 dex ($\sim$20$\times$) and falls to 0.9 dex ($\sim$8$\times$) between {log($\mu$)} $=$ -0.5$-$1.0 and 1.5. 

Overall, we do not see as strong a correlation between the excess volume density of AGN and the mass ratio between the galaxy pairs as the separation between the pairs. This may be because our mass ratios are calculated from the stellar mass within the central 1 kpc of the galaxies instead of the total stellar mass of the galaxies. Therefore cannot conclude that the AGN volume density is independent of the mass ratio.

\section{Enhanced AGN Luminosity}\label{sec:oiiilum} 

The source of the powerful emission we observe from AGN is a result of material accreting onto the black hole. Unfortunately, direct measurements of the bolometric luminosity for AGN are difficult as they emit across several wavelength regimes and many of those wavelength regimes are obscured. In the optical regime, the \OIII\ luminosity is often used as a proxy for the bolometric luminosity \citep{Heckman14}. The \OIII\ line is useful as it is strong enough to be visible in most galaxies and it is linearly proportional to the bolometric luminosity. 

\begin{figure*}
\centering
\includegraphics[width=7in]{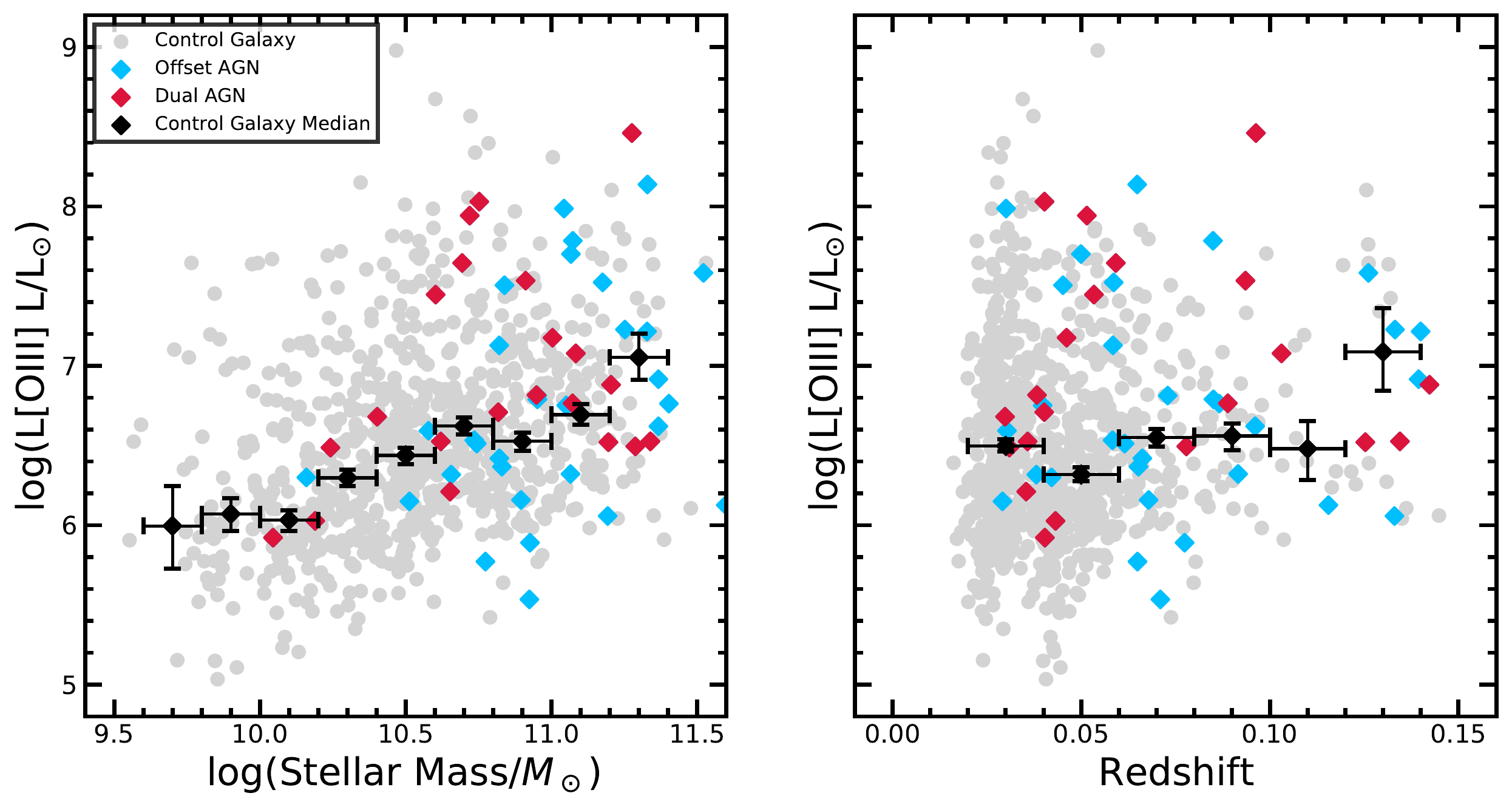}
\caption{Left: The \OIII\ Luminosity of the AGN in the pair and control samples as a function of the stellar mass. The grey circles are the AGN in the control galaxies, the blue diamonds represent offset AGN, and the red diamonds represent dAGN. In paired galaxies, only the MaNGA target is shown. The stellar masses for the paired galaxies are the total stellar masses from the NSA catalog and not the partitioned stellar masses described in Section \ref{sec:pair}. The black squares represent the median \OIII\ luminosity of the control sample within a discrete stellar mass bin. The horizontal bars represent the bin size and the vertical bars represent the 1$\sigma$ confidence intervals computed through bootstrap resampling. Right: The same figure as on the Left but for the \OIII\ luminosity of the AGN in the pair and control samples as a function of the redshift.}
\label{fig:oiii_ctrl}
\end{figure*}

Since the \OIII\ luminosity can be used as a tracer of the black hole accretion rate, we want to use this line to see if the black holes in paired galaxies accrete materials at higher rates than those found in isolated galaxies. We extract our luminosities from the central 1 kpc of the galaxies and correct our luminosities for reddening using the reddening curve from \citet{Calzetti00}, assuming $R_{\rm V}$ $=$ 3.1 and case-B recombination, H$\alpha$/H$\beta$ $=$ 2.86. As we had shown in Figure \ref{fig:agn_frac}, the fraction of AGN in our control sample is primarily a function of the galaxy's stellar mass. We plot the \OIII\ luminosity as a function of the stellar mass and redshift in Figure \ref{fig:oiii_ctrl}. We see that the \OIII\ luminosity increases with higher stellar masses and {the \OIII\ luminosity is almost constant with redshift up to $z$ $<$ 0.12.}

For a better comparison, we match each paired galaxy with control galaxies whose stellar masses are within 0.1 dex and whose redshift are within 0.01. We require that each paired galaxy have at least 20 control galaxies to match with and if they do not, they are excluded from the analysis.  We then take the difference between the luminosity of the paired galaxy and the median luminosity of its mass and redshift-matched controls. For this analysis, we only use the MaNGA target galaxy in each pair and we use total stellar masses from the NSA catalog for both the control galaxies and the paired galaxies. We decide to use the total stellar masses from the NSA catalog for the paired galaxies because the MaNGA survey has a tight stellar mass to redshift distribution. The paired galaxies whose stellar masses are split by our mass ratios often fall off of the survey's stellar mass$-$redshift distribution such that they find no similar control galaxies. 

\begin{figure}
\centering
\includegraphics[width=3in]{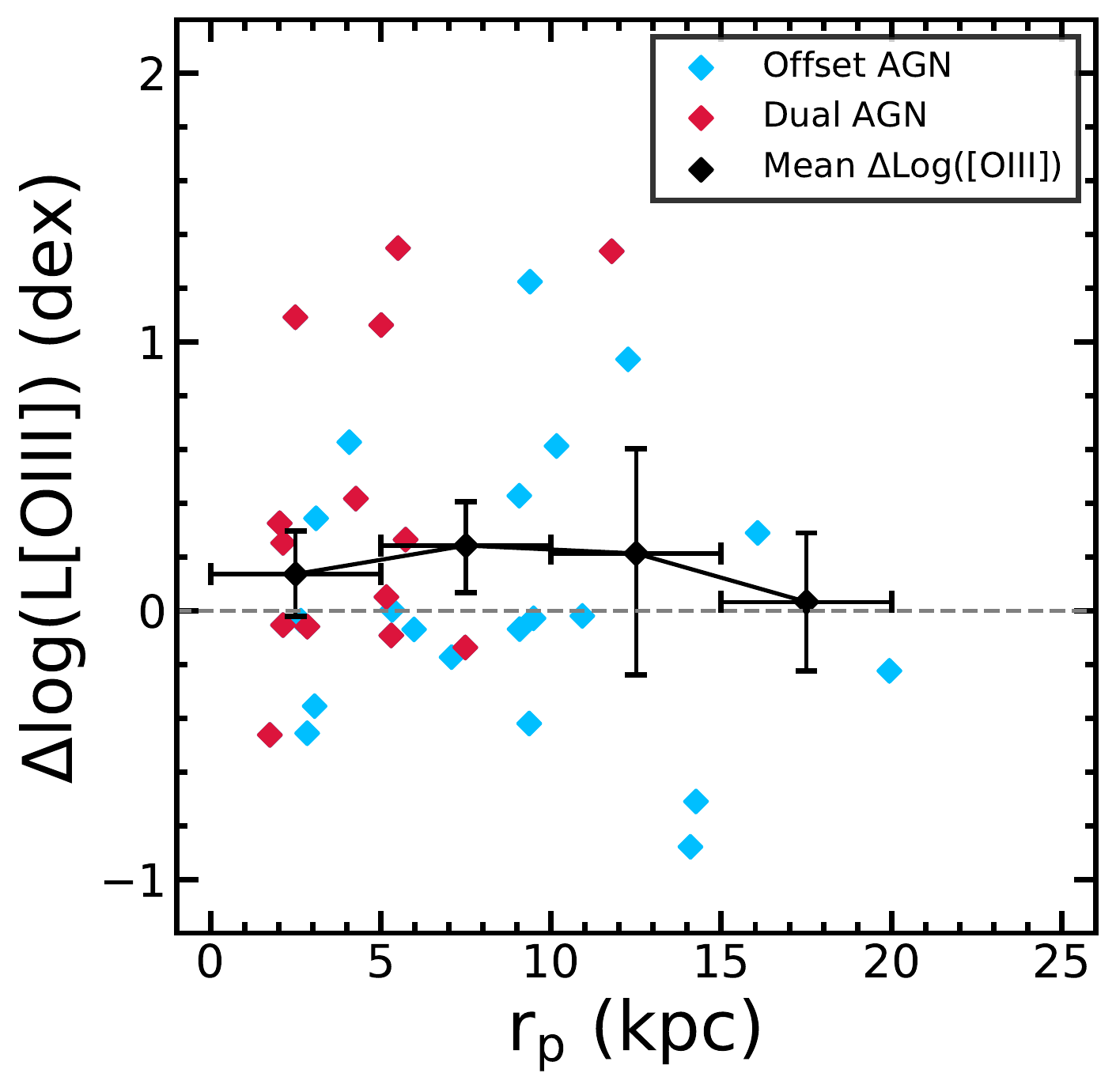}
\caption{The enhancement to the \OIII\ in the AGN in paired galaxies with respect to those in the control AGN. Blue diamonds represent the offset AGN sample and red diamonds represent the dAGN sample. The black line shows the {mean} $\Delta$log($L$\OIII) across different projected separation bins for the combined offset and dual AGN sample. The horizontal bars represent the bin size and the vertical bars represent the 1$\sigma$ confidence intervals computed through bootstrap resampling.}
\label{fig:oiii_diff}
\end{figure}

{In Figure \ref{fig:oiii_diff}, we present the excess \OIII\ luminosity of the AGN in paired galaxies with respect to the control sample.} We take the mean excess luminosity in projected separation bins, which are 5 kpc wide, and calculate the vertical error bars with the bootstrap resampling method mentioned in Section \ref{sec:den}.  Below 15 kpc, we see that the AGN in pairs have a \OIII\ luminosity that is 0.2 dex (1.6$\times$) higher than those in the control galaxies. This falls to zero beyond 15 kpc. Overall, we find that the \OIII\ luminosity of the paired galaxies is consistent with their control galaxies. About half of the AGN in pairs show enhanced \OIII\ luminosities, while the other half have lower luminosities than their controls. We do see that some paired galaxies have \OIII\ luminosities that are 10$-$20 times brighter than their control galaxies, though we find no correlation between these pairs and other merger parameters (like stellar mass, redshift, mass ratio, or the presence of tidal features). 

The relative \OIII\ luminosities have also been studied in previous works. \citet{Ellison13} found that the AGN in post-merger galaxies have \OIII\ luminosities that are 0.9 dex (8$\times$) higher than the AGN in controls and that the AGN in paired galaxies have \OIII\ luminosities that are 0.5 dex (3$\times$) higher than controls at separations of 5 kpc. This enhanced \OIII\ luminosity falls to 0.3 dex (2$\times$) between 20$-$25 kpc. The \OIII\ luminosity differences that we find are lower than those found in \citet{Ellison13}, by $\sim$0.3 dex. \citet{Jin21} also compares the surface brightness of \OIII\ between the AGN in pairs and the AGN in control galaxies using the MaNGA survey. They find that the \OIII\ surface brightness is 0.13 dex higher in the AGN in pairs than in the AGN in isolated galaxies which is consistent with what we find, although, due to the wide errors in our work, we conclude that the \OIII\ luminosity of the AGN in our paired galaxies is consistent with their control galaxies. Based on this, we infer that the black hole accretions rates between isolated and interacting galaxies are similar. This could mean that either merger-induced fueling occurs at a similar rate to stochastically induced fueling or that stochastic fueling is the dominant mode of AGN fueling even in our pair sample.

\section{Discussion}\label{sec:disc}

\subsection{Comparison to Paper I}
In this work we follow up on the preliminary AGN volume density from \citetalias{Fu18} and we find a greater excess of AGN in the pair sample than we did in our previous work in \citetalias{Fu18}. The work in \citetalias{Fu18} was done using an earlier MaNGA data release (DR14) in which only $\sim$2700 of the 10,000 MaNGA observations have been completed and found 105 galaxy pairs. In this work, we use MaNGA's final data release, containing around 10,000 observations. We find 391 pair systems which is what we would expect from a MaNGA sample that is 4$\times$ larger.

We use the same AGN selection in this work as we had in \citetalias{Fu18}, except that we use a stricter \ewha\ cut of 6\AA\ instead of the usual 3\AA\ cut. {In \citetalias{Fu18}, we found 391 AGN in the MaNGA survey with 50 AGN among the 105 galaxy pairs and a sample of 14 dAGN. In this work, we find {1045 AGN} in the MaNGA survey with 105 AGN among the 391 galaxy pairs and sample of 25 dAGN.} Despite the 4$\times$ larger sample size, we only find 2.6$\times$ the AGN as we had in \citetalias{Fu18}. This is due to the stricter \ewha\ cut that we employ in this work. {If we lower our \ewha\ cut to 3\AA, we find {1757 AGN} in the MaNGA survey with 154 AGN in galaxy pairs and sample of 44 dAGN, which is closer to what we anticipated from the complete MaNGA.} As we had mentioned in Section \ref{sec:control}, our models of the AGN fraction from Equation \ref{eq:mod} are consistent with each other if we assume a 3\AA\ cut to the \ewha\ in our AGN selection. This shows that while the absolute number of AGN in the sample did not linearly increase with the total number of galaxies observed by subsequent MaNGA data releases, the AGN fraction was preserved. 

\begin{figure*}
\centering
\includegraphics[width=6in]{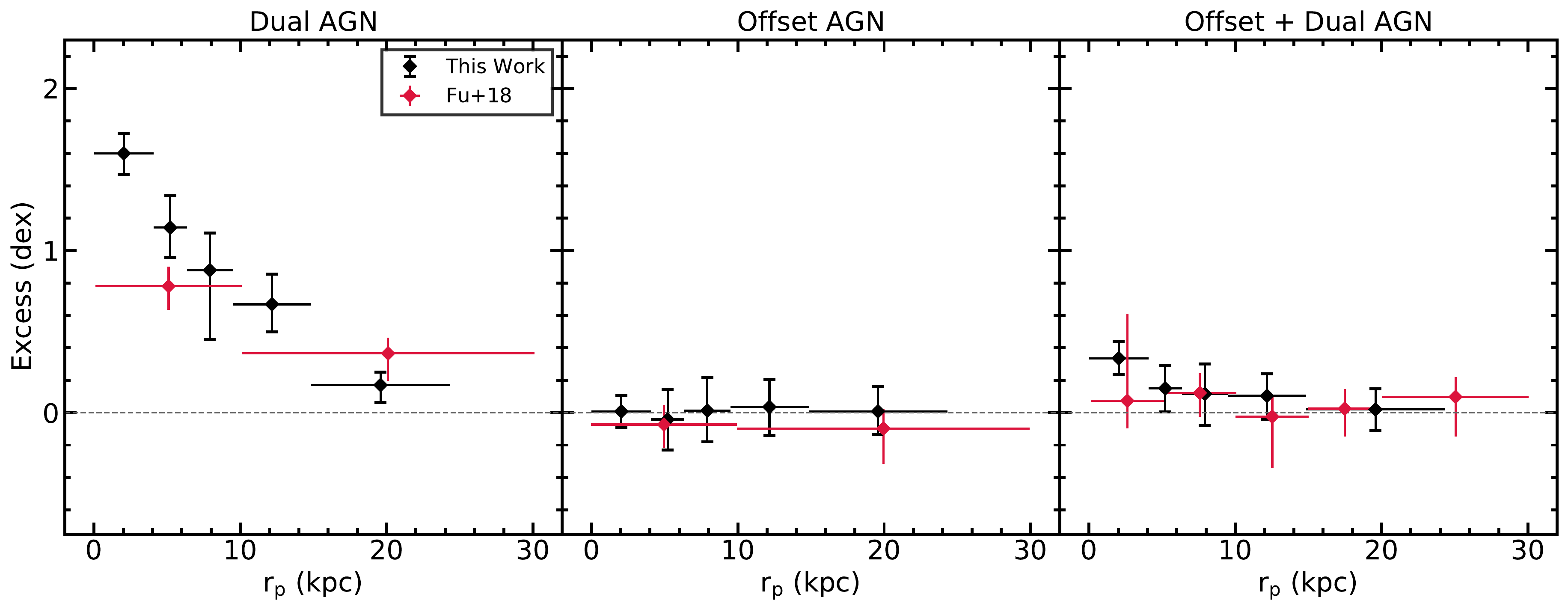}
\caption{The AGN excess in comparison to the random pairing of AGN from Figure \ref{fig:vol_den_sep} as a function of the projected separation. The figure is split into dAGN (left) offset AGN (middle) and offset+dual AGN (right). The results of this work (black diamonds) are shown alongside the previous results from \citetalias{Fu18} (red diamonds). {\citetalias{Fu18} had a smaller AGN sample and used a less strict \ewha\ cut (3\AA\ as opposed to 6\AA) to select AGN than this work.}}
\label{fig:fu18}
\end{figure*}

We compare the results of the work with \citetalias{Fu18} in Figure \ref{fig:fu18}. For offset AGN and offset+dual AGN, we find that our results are consistent with the results of \citetalias{Fu18}. For dAGN, our results are consistent beyond 6 kpc. Within 6 kpc, the excess dAGN is substantially higher than what we found in \citetalias{Fu18}. We find a dAGN excess of 1.6 dex (40$\times$) within 4 kpc while \citetalias{Fu18} found a dAGN excess of 0.8 dex (6.3 $\times$) within 10 kpc. 

The differences that we observe between the two works are likely due to the larger sample size from the MaNGA survey's final data release. The large sample of galaxy pairs and AGN gave us better statistics for our volume density calculations. This has allowed us to use smaller bin sizes in Figure \ref{fig:fu18} which has unveiled substantial excess of dAGN and a minor excess of offset+dual AGN under a separation of 4 kpc. The minor excess of offset+dual AGN that we detect in this work (and that we couldn't detect in \citetalias{Fu18}) is significant because it allowed us to break the degeneracy between the effects of merger-enhanced AGN fueling and correlated AGN activity.

\citetalias{Fu18} estimates the rate of AGN correlation, $\xi$, based on the excess of dAGN. \citetalias{Fu18} estimates that $\xi$ is 15\% at separations of 20 kpc and 40\% at 5 kpc. In our work, Equation \ref{eq:xi} predicts a rate of correlation that is 6\% at 20 and kpc 20\% at 5 kpc. The rates of correlation in this work are lower because we include a second term in our model, the merger-induced fueling. The result of this is that the excess of dAGN is split between correlated activity and merger-induced fueling. We are able to include the merger-induced fueling term in this work because we now observe an excess AGN in the offset+dual AGN sample where we did not observe this excess in the previous work. 

\subsection{Comparison with Other Observations}
\begin{figure}
\centering
\includegraphics[width=3in]{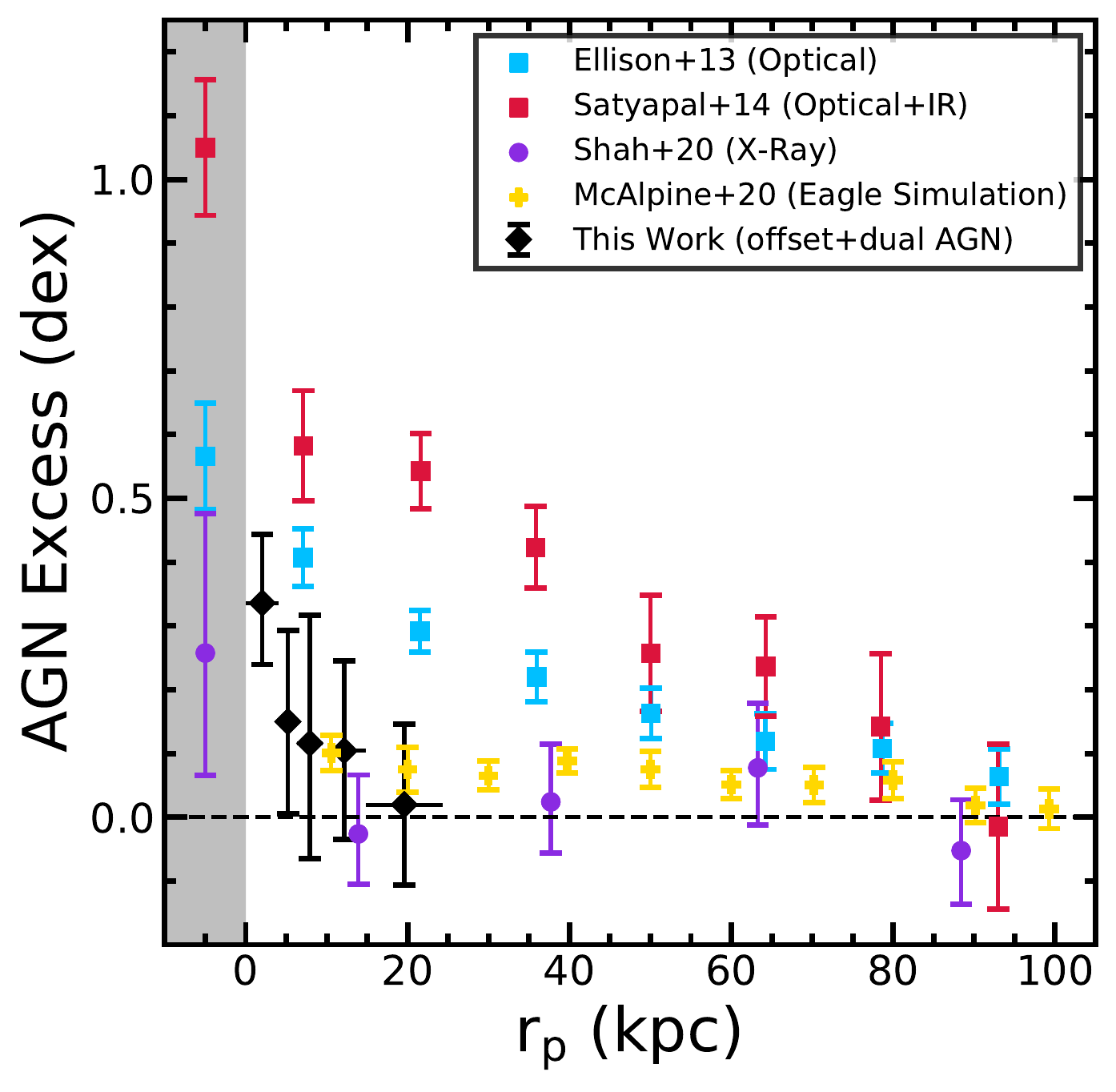}
\caption{We show the AGN excess from this work (black) and other previous works; \citet{Ellison13} (blue), \citet{Satyapal14} (red), \citet{Shah20} (purple), and \citet{McAlpine20} (yellow). The data points below a projected separation of 0 kpc represent a sample of post-merger galaxies.}
\label{fig:compare}
\end{figure}

In this section we compare the results of this work against other previous works which find an excess of AGN in galaxy pairs. Figure \ref{fig:compare} shows the AGN excess of the offset+dual AGN sample from our work and the optically selected AGN from \citet{Ellison13}, the optical and infrared selected AGN from \citet{Satyapal14}, the X-ray selected AGN from \citet{Shah20}, and the AGN excess in pairs predicted by the Eagle simulation in \citet{McAlpine20}. 

The pair sample in \citet{Ellison13} consisted of $\sim$1700 galaxy pairs and 97 post-merger galaxies from SDSS's spectroscopic survey. The galaxy pairs were selected to have a projected separation of 80 $h^{-1}_{70}$ kpc, line-of-sight velocity difference of 300 km s$^{-1}$, a mass ratio within 4:1 stellar masses, and a redshift between 0.01$\le$ $z$ $\le$ 0.2.  The post-merger galaxies were visually identified by their morphology with Galaxy Zoo's citizen science project. AGN are identified though BPT analysis using \citet{Stasinska06} as the cut between AGN and star-forming galaxies. AGN excesses are then calculated by matching each paired galaxy with a set of control galaxies which have similar stellar masses, redshifts, and environmental densities. 

\citet{Ellison13} found that the AGN excess in paired galaxies is $\sim$2.5$\times$ higher than controls at a projected separation of 10 kpc. This excess falls to one at a separation of 90 kpc. This AGN excess is roughly consistent with ours at close separations; however, our excess falls to one at a much closer separation, only 20 kpc. The differences between our two samples may be due to differences in our sample selections. \citet{Ellison13} uses a more relaxed AGN selection than ours, using the cut from \citet{Stasinska06} while we use the AGN cut from K03. Further, we use a \ewha\ cut of 6\AA\ to remove retired galaxies from our AGN sample while \citet{Ellison13} uses a signal-to-noise ratio cut of 5$\sigma$ on the emission line fluxes needed for the BPT analysis to remove retired galaxies. 

\citet{Satyapal14} used the same pair sample as \citet{Ellison13}; however, they used infrared color cuts with WISE along with optical BPT cuts to select AGN. The excess AGN seen in \citet{Satyapal14} is roughly twice the excess seen in \citet{Ellison13} for all separations. The large excess of AGN seen in \citet{Satyapal14} is likely due to the usage of infrared to select AGN. Infrared selection can identify AGN that have been optically obscured by dust. The tidal disruptions from merger events can dredge up large amounts of dust which can optically obscure AGN in their cores. \citet{Weston17} suggests that as many as 50\% of the AGN in galaxy pairs may be optically obscured by dust, which corresponds to \citet{Satyapal14} finding an AGN excess that is almost twice are large as the one in \citet{Ellison13}. 

\citet{Shah20} built a sample of 2381 spectroscopic pairs from the CANDELS \citep{Grogin11,Koekemoer11} and COSMOS \citep{Scoville07} surveys. These pairs had a redshift range between 0.5$-$3, a stellar mass greater than 10$^{10}$ M$_{\odot}$, a mass ratio within 4:1, a relative line-of-sight velocity within 1000 km s$^{-1}$, and a projected separation within 150 kpc. This sample is supplemented with a set of pairs with visible tidal features and post-merger galaxies from \citet{Kartaltepe15}. AGN were then selected to have an X-ray luminosity above 10$^{42}$ erg s$^{-1}$ with deep {\it Chandra} X-ray observations. AGN excess were then calculated by matching paired galaxies to a set of 3 control galaxies with similar stellar masses, redshifts, and environmental densities.

\citet{Shah20} found no enhancement to X-ray selected pairs at all separations, but their sample of visually identified pairs/post-mergers do feature excess AGN. The excess AGN in the visually identified pairs/post-mergers is consistent with our results and is about 0.1 dex lower than the results that we find in our closest pairs. The AGN excess is lower than both the AGN excess seen in the post-merger samples of \citet{Ellison13} (Optical) and \citet{Satyapal14} (Optical+IR). \citet{Shah20} suggested that the low excess of AGN that they observe may be due to the high redshift range of their sample. 

{\citet{Jin21} builds a sample of paired and post-merger galaxies from the MaNGA Product Launch-6 and SDSS DR15 (4691 galaxies; \citet{Aguado19}). The sample includes 707 paired galaxies that are found by crossmatching the MaNGA target galaxies with galaxies in the NSA catalog. They further find another 46 pairs in which each galaxy is covered by a separate MaNGA IFU. This selection method misses late-stage mergers whose companions lack redshifts and coalescing mergers which appear to be a single galaxy. To recover these, \citet{Jin21} visually inspects the MaNGA fields for mergers based on their morphologies in SDSS {\it gri} images. This adds an additional 116 paired galaxies and 125 post-merger galaxies. 

\citet{Jin21} studies the rate of AGN in pairs as a function of merger sequence. The study found no evolution of the AGN fraction across different merger sequences and found no difference between the rate of AGN in galaxy pairs and AGN in isolated galaxies. The differences between our works may be due to a couple of reasons. First, the apertures from which we are extracting the spectra for our galaxies are different. \citet{Jin21} selects the inner 3$\times$3 spaxels of the galaxy corresponding to a 1.5\arcsec$\times$1.5\arcsec\ square on the sky. Since these apertures have a fixed angular size, the apertures will cover a larger physical area on the higher redshift than on the lower redshift galaxies. Because of this, there may be cases in which the core of the galaxy is only contained within a minority of the selected spaxels and that the majority of the spaxels are covering off-nuclear regions. Since AGN activity is typically concentrated in the centers of galaxies, this will result in cases in which AGN spectra from the core of a galaxy is being diluted by non-AGN spectra from off-nuclear regions. As a result, some AGN may be missed by the BPT analysis. Our survey lessens the impact of this issue by implementing physical 1 kpc radius circular apertures. The second difference is that our AGN fraction is corrected for volume, while \citet{Jin21} does not apply a volume correction. Given the significantly different sample selection methods and the application of volume correction, it is hard to compare directly our increasing AGN fraction trend with the flat fraction in various merger stages found in \citep{Jin21}. 

\citet{Stemo21} builds a catalog of 204 AGN in mergers observed by the Hubble Space Telescope between a redshift range of 0.2 $<$ $z$ $<$ 2.5. They find that AGN activation increases steeply at close bulge separations, $<$4 kpc and at bulge separations of 12$-$14 kpc. These two peaks of AGN activity are likely reflective of the galaxies passing their first and second pericenters during the merger event. In our work, we do not see a substantial secondary peak around 12$-$14 kpc; however, our catalog does not feature as many AGN at that separation range.

We find that our AGN excess is in good agreement with previous measurements of the AGN excess in paired galaxies at short separations, {at least for galaxy pairs in the local universe}. The AGN excess that we observe is generally lower in comparison to previous works and the AGN excess falls more rapidly with increasing separation. One potential problem is that our sample will miss galaxy pairs in which one galaxy is in the MaNGA IFU and the another is just outside of the IFU. This means that our pair sample is incomplete at certain separations depending on the IFU size. On the other hand, by using this pair selection that is entirely contained in the MaNGA survey, we can control for the well known biases of the MaNGA survey. Our pair sample also contains no post-merger galaxies since our pair selection method requires two distinct galaxy cores. 

\subsection{Comparison to Simulations}

\citet{McAlpine20} studies the AGN fraction in pairs with the {\sc eagle} simulation (Evolution and Assembly of GaLaxies and their Environment; \citet{Crain15}; \citet{Schaye15}). The AGN are selected to have bolometric luminosities greater than $L_{\rm bol}$ $\ge$ 10$^{43}$ erg s$^{-1}$ and Eddington rates greater than $\lambda_{\rm edd}$ $\ge$ 10$^{-2}$. Control galaxies are selected to have similar redshifts, stellar masses, halo masses, gas masses, and black hole masses as the paired galaxies that they are matched to. In Figure \ref{fig:compare}, we show the AGN excess from \citet{McAlpine20} in major mergers with redshifts under $z$ $\le$ 1.0. The AGN excess is weaker than those found \citet{Ellison13} and \citet{Satyapal14} at all separations. At 10 kpc, \citet{McAlpine20} finds a AGN excess of $\sim$1.3$\times$ which is similar to, but slightly smaller than, the excess we find at the same separation ($\sim$1.6$\times$). 

\citet{Steinborn16} studies the properties of AGN in galaxy pairs with the cosmological hydrodynamical simulations from the Magneticum Pathfinder Simulation set \citep{Dolag16}. The work includes 34 galaxy pairs at a $z$ $=$ 2. Within these pairs there are 9 dAGN, 14 offset AGN, and 11 pairs with no AGN activity. \citet{Steinborn16} found that dAGN tend to exist in pairs with small separations, $<$5 kpc, and that the pairs without AGN activity tend to have wider separations. We likewise observe the largest excess of dAGN below 5 kpc.

There are a few key differences between observational studies and simulations like \citet{Steinborn16} and \citet{McAlpine20}. First, the separations in the simulations are real 3D separations and not 2D projected separations that assume that the two galaxies are at the same distance along our line-of-sight. Second, the merger simulations can tell if the paired galaxies have or have not interacted yet. With observations, we only have a snapshot of the merger event. We do not know if the two galaxies are on their first approach (and have not interacted yet) or if they are between their first or second pericenter. Third, simulations can tell if a galaxy hosts an AGN without observational effects like dust obscuration. Despite these differences, our results are in good agreement with those from \citet{Steinborn16} and \citet{McAlpine20}.

\section{Conclusion}\label{sec:conl}

In this work, we identified a sample of 391 galaxy pairs within the fields-of-view of {8,585 MaNGA IFUs.} We identified 105 AGN in the pair sample {using optical BPT analysis}. Among these, we found 25 dAGN systems. We found that galaxy pairs have a greater fraction of AGN than control galaxies in the MaNGA survey. We summarize the findings of this work below. 

\begin{enumerate}
	\item We found that the AGN in galaxy pairs have a volume density that is 2.5$\times$ higher than what would be expected by random pairing at close separations. This excess of AGN disappears around 20 kpc. In dAGN, we found a volume density of AGN that is 40$\times$ higher than what would have been expected by random pairing at close separations. This excess of AGN falls to 1.6$\times$ at separations of 20 kpc. 
	
	\item We model the observed AGN volume density in galaxy pairs to account to stochastic AGN fueling, merger-enhanced fueling, and correlated AGN activity. We demonstrate that excess AGN are induced in mergers not only from synchronous AGN activity, but also that the merger event can increase the natural rate of AGN fueling.
	
	\item Our model predicts that 11.5\% of offset AGN will be converted to dAGN at a projected separation of 10 kpc. Our model further predicts that merger-induced fueling will enhance the rate of stochastic fueling by a factor of 1.27$\times$ at 10 kpc. 

	\item We found that the \OIII\ luminosity of the AGN in paired galaxies is consistent with the AGN in mass and redshift matched control galaxies. This demonstrates that galaxy interactions may increase the likelihood of activating an AGN, but they do not necessarily induce more luminous AGN than stochastically induced AGN. 

\end{enumerate}

In this work, {we confirm our results obtained in} \citepalias{Fu18} in that we have been able to demonstrate that galaxy mergers are able to induce AGN activity in the centers of galaxies. {We also show that the \OIII\ luminosities of the AGN induced by galaxy interactions are consistent with the AGN in isolated galaxies. This indicates that stochastic fueling may still be the dominant fueling mechanism in close pairs. This also demonstrates how it is difficult to separate effects of tidally induced and stochastically gas-inflows in galaxy pairs. } 



We thank the anonymous referee for useful comments that helped improve the manuscript. J.S. and H.F. acknowledge support from the National Science Foundation (NSF) grants; AST-1614326 and AST-2103251. YSD  acknowledges the support from the National Key R\&D Program of China via grant No.2022YFA1605000, and the National Natural Science Foundation of China grants 12273051 and 11933003. ICG would like to acknowledge support from DGAPA-UNAM grant IN113320. Funding for the Sloan Digital Sky Survey IV has been provided by the Alfred P. Sloan Foundation, the U.S. Department of Energy Office of Science, and the Participating Institutions. SDSS acknowledges support and resources from the Center for High-Performance Computing at the University of Utah. The SDSS web site is \url{www.sdss.org}.

SDSS is managed by the Astrophysical Research Consortium for the Participating Institutions of the SDSS Collaboration including the Brazilian Participation Group, the Carnegie Institution for Science, Carnegie Mellon University, Center for Astrophysics | Harvard \& Smithsonian (CfA), the Chilean Participation Group, the French Participation Group, Instituto de Astrofísica de Canarias, The Johns Hopkins University, Kavli Institute for the Physics and Mathematics of the Universe (IPMU) / University of Tokyo, the Korean Participation Group, Lawrence Berkeley National Laboratory, Leibniz Institut für Astrophysik Potsdam (AIP), Max-Planck-Institut für Astronomie (MPIA Heidelberg), Max-Planck-Institut für Astrophysik (MPA Garching), Max-Planck-Institut für Extraterrestrische Physik (MPE), National Astronomical Observatories of China, New Mexico State University, New York University, University of Notre Dame, Observatório Nacional / MCTI, The Ohio State University, Pennsylvania State University, Shanghai Astronomical Observatory, United Kingdom Participation Group, Universidad Nacional Autónoma de México, University of Arizona, University of Colorado Boulder, University of Oxford, University of Portsmouth, University of Utah, University of Virginia, University of Washington, University of Wisconsin, Vanderbilt University, and Yale University.

\bibliography{bib}
\bibliographystyle{aasjournal}

\end{document}